\documentclass[superscriptaddress,longbibliography, reprint,amsmath,amssymb,aps,pra,floatfix]{revtex4-2}

\usepackage{comment}

\usepackage{subfigure}
\usepackage{multirow}
\usepackage{fancyhdr}
\usepackage{longtable}
\usepackage{parskip}
\usepackage[T1]{fontenc}
\usepackage{dcolumn}   % Align table columns on decimal point
\usepackage{physics} % for \ket and other quantum notation
\usepackage{bm}        % bold math
\usepackage{amsfonts}  % Common math fonts
\usepackage{amsmath}   % Common math functions
\usepackage{amssymb}   % Common math symbols
\usepackage{xcolor}
\usepackage{multirow}

\usepackage{float}

\usepackage{graphicx}

\usepackage{hyperref}
\hypersetup{
    colorlinks=true,
    linkcolor=blue,
    citecolor=magenta,
    bookmarksdepth=5
}

\begin{document}

\title{Passive quantum error correction of photon loss at breakeven}

\author{Shruti Shirol, Sean van Geldern, Hanzhe Xi, Chen Wang}

\affiliation {\it Department of Physics, University of Massachusetts-Amherst, Amherst, MA 01003}

\date{October 22, 2025}
%\end{comment}

\begin{abstract} %\bfseries \boldmath
Physical qubits in a quantum computer are often represented by superposition states of single particles or excitations. Decay of the excitation itself is a fundamental error channel that is difficult to overcome via external drive or control techniques. Quantum error correcting codes, which encode information in superpositions involving multiple excitations, provide a path to preserve information beyond the capacity of individual excitations, 
but typically require exquisite active operations on the system.  Here, we demonstrate a steady-state driven dissipative quantum system, composed of a superconducting cavity and a transmon ancilla, that preserves a logical qubit beyond the photon-lifetime limit by about 5\% using a binomial encoding. %encoding of photon number states.  
This realization of continuous quantum error correction at the breakeven point highlights the quantitative competitiveness of passive correction strategies while circumventing some demanding hardware requirements of its active counterparts. 
%the benefits of dissipation engineering for hardware-efficient passive protection of quantum information. 

\end{abstract}
%\pacs{47.15.-x}

\maketitle 
\noindent
%\section*{Introduction}
Environmental noise in quantum computing devices leads to stochastic evolution of quantum states, i.e.~decoherence, the effects of which must be suppressed to avoid logical errors in the computation.  Strategies to counter decoherence have been jointly pursued on a stack of three levels: construction of the physical device, choices of the encoding and external drive fields, and implementations of quantum error correction (QEC) circuits. Notably in the middle level, suppressing errors passively by continuously driving a quantum system~\cite{timoney2011quantum, CDDinDiamond, leghtas2015confining, grimm2020stabilization, %mundada2020floquet, reglade2024quantum
laucht2017dressed} offers strong appeal in delivering performance gains with minimal overhead.  The possibility of such passive protection depends on the nature of the physical noise channels.  A common class of noise manifests as quasi-static fluctuations of Hamiltonian parameters, e.g.,~$1/f$ dephasing~\cite{paladino20141}, which can be suppressed by continuous dynamical decoupling without incurring quantum resource overhead~\cite{viola2003robust, yan2013rotating, huang2021engineering, nguyen2024programmable%,  passiveDDTrappedIon, SCqubitsRobustSpinLocked
}. Suppression of dephasing up to higher order has further been accomplished by incorporating QEC encodings in the expanded Hilbert space of driven and/or dissipative oscillators~\cite{hajr2024high, leghtas2015confining, %lescanne2020exponential, 
reglade2024quantum, grimm2020stabilization, frattini2024observation} and high-spin systems~\cite{li2025beating,debry2025errorcorrectionlogicalqubit}.  

However, some decoherence processes are practically broadband %(i.e.~discrete quantum jump-like) 
and necessitate the use of QEC codes to protect the information even at the lowest order. Stochastic loss of elementary excitations, such as $T_1$ decay of qubit excitations or cavity photons, belongs to this category, and is the current limiting factor of superconducting quantum processors. 
To overcome this limit, active QEC has been a focus of pursuit using e.g.~the surface code for transmon qubits~\cite{acharya2024quantum, krinner2022realizing}, the repetition code for dissipative cat qubits~\cite{putterman2025hardware}, and a variety of bosonic codes~\cite{campagne2020quantum,  hu2019quantum, ofek2016extending,sivak2023real,brock2024quantumerrorcorrectionqudits, ni2023beating}. 
Due to the quality requirements of the system components and control infrastructure to implement active QEC, only a few state-of-the-art experiments on $T_1$-limited quantum hardware are able to overcome the QEC overhead to reach the so-called break-even point~\cite{ofek2016extending, ni2023beating, sivak2023real, acharya2024quantum, brock2024quantumerrorcorrectionqudits}, where the error-corrected logical qubit retains information longer than any individual physical component of the system. %time of the underlying individual physical excitations.  %In the quest for more efficient protection of logical qubits, there is growing interest in implementing QEC more autonomously by cyclic applications of unitary gates and unconditional reset without the need for measurement feedback~\cite{}

Is it possible to engineer continuous quantum dynamics to protect logical information beyond the lifetime limit of the information-carrying excitations?  The foundations of autonomous QEC (AQEC) have been explored since the early 2000's~\cite{AhnDoherty2013book, ahn2002continuous}
%sarovar2005continuous, beige2000quantum, 
where continuously-acting Markovian dissipation is considered for reverting the action of discrete errors %such as excitation losses 
as they stochastically occur.  Several practical implementations of continuous AQEC in superconducting devices have been proposed~\cite{cohen2014aqecsccircuits, albert2019pair, xu2023autonomous, kapit2016aqecsccircuits} and implemented~\cite{gertler2021protecting, li2024autonomous, livingston2022contqec}. However, no continuous AQEC experiments so far are able to tailor the dissipation sufficiently well to %fully overcome the QEC overhead and 
reach the break-even point. 

In this work, we demonstrate continuous and autonomous quantum error correction at the break-even point for a superconducting cavity qubit.  The logical qubit is encoded with a binomial bosonic code and passively protected from single photon losses by an efficient parity-recovery dissipation enabled by a three-level superconducting transmon ancilla.  We report a time-averaged logical qubit coherence time of $196\pm1$ $\mu$s that narrowly exceeds the $T_1$-limited coherence time of individual photons in the cavity, $186\pm1$ $\mu$s.  %Our experiment therefore shows that, under continuous dissipation, the passive information storage lifetime of a quantum system can exceed the physical lifetime limits of its constituent excitations, suggesting an efficient strategy to protect logical qubits at the hardware level.  
We further discuss the resemblance of the continuous AQEC actions to active QEC cycles, as well as their corresponding remaining error channels.  %Our analysis shows that the passive method is capable of delivering comparable and potentially superior performance compared to active QEC in protecting a bosonic cavity memory in circuit QED.  
Together with a growing frontier of discrete-cycle-based AQEC strategies~\cite{lachance2024autonomous, de2022error} (which have also reported break-even performance during the preparation of this manuscript~\cite{sun2025extending, ni2025autonomous} using similar binomial encodings), our results add to an emerging array of strategies to extend logical coherence times at the interface of quantum hardware engineering and error correction.

\section*{Passive QEC Technique}

Our AQEC scheme targets a class of rotation-symmetric bosonic codes~\cite{grimsmo2020rsbc} that encode a logical qubit in an eigenspace of the generalized photon-number parity operator of a cavity, $\hat{\Pi}_m=e^{i\frac{2\pi}{m}\hat{a}^\dagger\hat{a}}$, using code words %corresponding to the photon-number modulo $m$ 
with equal mean photon number $\bar{n}$. For $m=2$, %we get the minimal parity syndrome allowing 
this construction divides the Hilbert space into even and odd photon-number subspaces, sufficient to correct a single-photon loss event.  Prominent examples of experimental implementation include the 4-legged cat code~\cite{ofek2016extending} and the kitten code~\cite{ni2023beating, hu2019quantum}, where repeated measurements of the photon number parity led to real-time detection of photon-loss errors and recovery of the logical information.  

\begin{figure} 
	\centering
	\includegraphics[width=8.6cm]{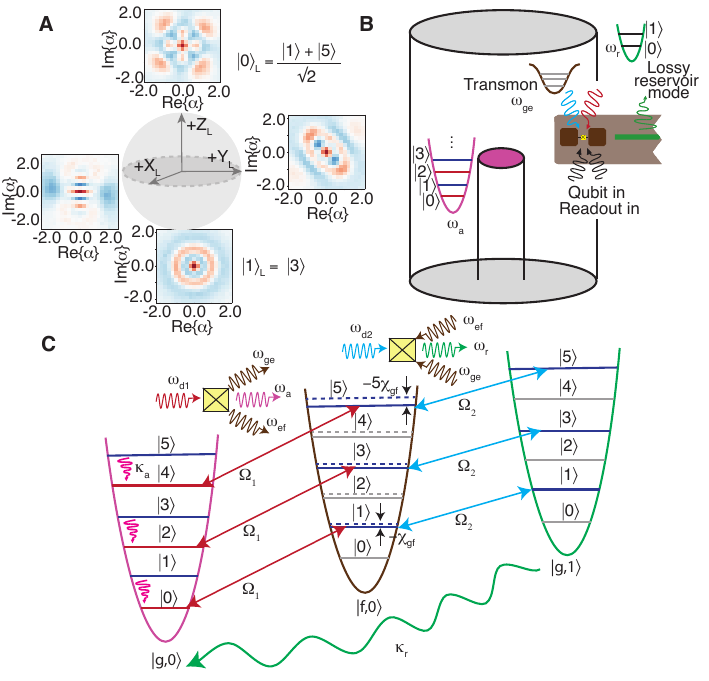}

	\caption{\textbf{Experimental setup and correction protocol}
		 (\textbf{A}) The Bloch sphere representation of the logical qubit, and the measured Wigner function of the pole ($|0_{L}\rangle$, $|1_{L}\rangle$) and the equator ($|+X_{L}\rangle$,  $|+Y_{L}\rangle$) states. 
         %The states are prepared in the storage cavity using optimal control theory (OCT) pulses. 
         (\textbf{B}) Cartoon depiction of the experimental device, showing the 3D post cavity storage resonator, transmon ancilla and quasi-planar reservoir/readout mode. One input port is used for transmon drive, transmon readout and the AQEC drives as shown. 
         %There is a separate drive port to displace the storage cavity mode.
         (\textbf{C}) Level diagram depicting the transition paths of the AQEC scheme, implemented with two combs of simultaneous continuous drives (red arrows and blue arrows) and dissipation from the reservoir mode (green squiggly arrows).  Above the drive-tone arrows, four-wave-mixing diagrams denote the conversion of excitations for each drive. 
         }
	\label{fig:fig1} % give each figure a logical label name
\end{figure}

In contrast, our scheme passively restores the correct parity by engineering a continuous dissipative process, eliminating the need for real-time feedback.  Specifically, we implement Parity Recovery by Selective Photon Addition (PReSPA)~\cite{gertler2021protecting} that adds a photon only when the storage mode is in the even-parity error subspace. It can be described by an effective Linbladian dissipator $\mathcal{D}[\hat{\Pi}_{cor}]$, where
%\begin{equation}
$\hat{\Pi}_{cor}=\sqrt{\kappa_{cor}}\sum_n|2n+1\rangle\langle2n|$.
%\end{equation} 
The PReSPA operator realizes an approximate recovery map that fully preserves the logical information in the limit of $\bar{n}\rightarrow\infty$.  At finite photon numbers, distortion of the codeword coefficients over time leads to a secondary logical dephasing~\cite{gertler2021protecting}, which may be corrected with an engineered 4-photon dissipation in the future~\cite{mirrahimi2014dynamically, vanselow2025dissipatingquartets}.

In this demonstration, we employ a binomial code in an odd-parity subspace with $\bar{n}=3$, 
\begin{equation}
	|0_{L}\rangle = \frac{|1\rangle + |5\rangle}{\sqrt{2}}, 
    |1_{L}\rangle = |3\rangle
	\label{eq:log_state} % Use a logical label
\end{equation}
which is illustrated in Fig.~\ref{fig:fig1}A with its logical Bloch sphere representations.  The choice is a result of balancing the codeword distortion effect at small $\bar{n}$ and the increased rate of photon loss at larger $\bar{n}$.

To experimentally implement the dissipation operator $\hat{\Pi}_{cor}$ in a storage cavity mode, we use two auxiliary quantum modes, as represented in Fig.~\ref{fig:fig1}B. The nonlinear transmon ancilla allows photon-number-selective coherent control of the storage mode through the dispersive shift ($\chi_{ge}=1.12$ MHz and $\chi_{gf}=2.07$ MHz).  The quasi-planar reservoir mode \cite{axline2016architecture} with a short lifetime ($\kappa_r/2\pi=0.58$ MHz) extracts entropy by dissipating photons. The relevant terms of system Hamiltonian %with all the couplings between modes 
in the lab frame is, 
\begin{align} 
	\frac{\hat{H}_0}{\hbar} = &\omega_{a}\hat{a}^{\dagger}\hat{a} + \omega_{q}\hat{q}^{\dagger}\hat{q} + \omega_{r}\hat{r}^{\dagger}\hat{r} - \frac{\alpha_{q}}{2}\hat{q}^{\dagger2}\hat{q}^2  -\chi_{ge}|e \rangle\langle e| \hat{a}^{\dagger} \hat{a} \nonumber \\
    &\quad -\chi_{gf}|f \rangle\langle f|\hat{a}^{\dagger}\hat{a} - \chi_{qr} |e \rangle\langle e| \hat{r}^{\dagger}\hat{r}
    \label{eq:ham_1}
  % Use a logical label
\end{align}
\noindent where $\hat{a}, \hat{q}, \hat{r}$ are the bosonic annihilation operators of the storage, transmon, and reservoir modes, respectively. $\ket{e}$ and $\ket{f}$ represent the first and second excited state of the transmon, and $\alpha_q$ is the transmon anharmonicity. Other higher-order terms (such as cavity anharmonicity) have negligible effects to the AQEC process and are not included in Eq.~(\ref{eq:ham_1}). More details of the Hamiltonian parameters and the coherence times of the components are discussed in the supplementary Table \ref{tab:system_params}. 

Two combs of continuous drives at frequency $\omega_{d1}=\omega_a+(2\omega_q - \alpha-n\chi_{gf})$ and $\omega_{d2}=\omega_r- (2\omega_q-\alpha -n\chi_{gf})$ for $n=1,3,5$ are applied simultaneously which implement an effective two-stage cascaded dissipative process, shown in Fig.~\ref{fig:fig1}C. The drive at $\omega_{d1}$ activates the parity-selective photon-addition and the drive at $\omega_{d2}$ implements ancilla reset by converting the transmon excitation to a reservoir mode photon that is spontaneously emitted. These drive tones realize a rotating-frame (rotating with $\hat{H}_0$) Hamiltonian under rotating-wave approximation: 
\begin{align}
    \label{eq:ham_drive}
    \frac{\hat{H}_{d}}{\hbar} =&  \sum_{n=1,3,5}\Omega_{1} |n,f,0\rangle \langle n-1, g, 0| \nonumber \\ &+\Omega_{2} |n,g,1\rangle \langle n, f, 0 | + h.c.
\end{align}

The matched values of $\omega_{d1}+\omega_{d2}$, $\Omega_1$, and $\Omega_2$ across all three even-to-odd Fock-state conversion processes ensure that the frequency and the temporal profile of the emitted photon from the reservoir mode are identical, which conceals the which-path information and preserves the coherence of PReSPA. Similar comb driving techniques targeting a specific set of photon number states have also been recently applied in the context of parity mapping and unitary cavity control~\cite{ni2023beating, kudra2025photonadd}. 

To obtain fast error correction with minimal spurious actions on the code states, there are two key considerations in our implementation of PReSPA.  First, we exploit the intrinsically strongest four-wave mixing (FWM) sideband transitions available in the transmon-cavity system, which use the $|f\rangle$-state of the transmon as the intermediate state \cite{huang2025fast}.  Compared to the previous study that transits through the $\ket{e}$-state of the transmon \cite{gertler2021protecting}, this strategy allows us to obtain larger drive rate $\Omega_2$ at much weaker drive fields and hence poses much lower risk of spurious transitions.  The use of $\ket{f}$-state also benefits from its larger dispersive shift, allowing for the use of larger $\Omega_1$, thus faster overall correction, while satisfying the parity selectivity requirement $\Omega_1\ll\chi_{gf}$.

Second, we choose a rate hierarchy near the critical damping condition that yields the fastest non-oscillatory relaxation of the logical state back to the code space.  We find this condition to be $\Omega_1:\Omega_2:\kappa_r\approx\frac{1}{13}:\frac{1}{4}:1$ in a two-stage cascaded ``optical pumping'' model of PReSPA, which is a rather generic extension of the well-known $\Lambda$-system model. As we discuss in the Supplementary text, the effective decay rate of the initial state can be inferred from the imaginary component of the eigenvalues of a 3$\times$3 effective non-Hermitian Hamiltonian, as shown in Fig.~2B (which shows some surprisingly rich behavior despite the simplicity of the model).  Compared to operating in the adiabatic regime with strong $\kappa_r$, this strategy allows us to obtain the fastest possible effective dissipation given any practical constraints on $\Omega_1$ or $\Omega_2$.

\begin{figure*}[tbp] % Do NOT use \begin{figure*}
	\centering
	\includegraphics[width=\textwidth]{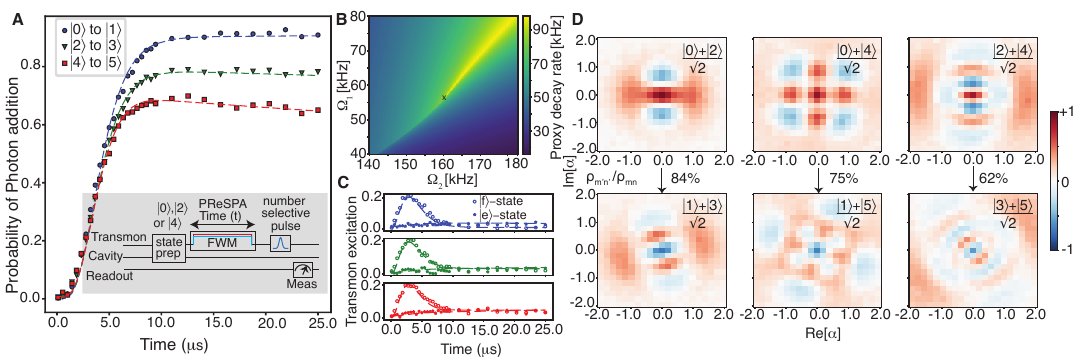} 
	
	\caption{\textbf{Characterization of PReSPA performance on even-odd conversion.}
		(\textbf{A}) Starting in even-parity Fock states, we show the cavity populations in converted odd-parity Fock states over pump times. % to find the rates and conversion efficiencies of the dissipative photon addition. 
        The cavity population in $\ket{n}$ presented here corresponds to $P_{\ket{n,g}}-P_{\ket{n,e}}$, hence a conservative estimate of the success probability of photon addition. (See Supplementary text for discussions.)
        Dashed lines: fit to the master-equation simulations, with $\Omega_{1} = 55$ kHz and $\Omega_{2} = 160$ kHz. (\textbf{B}) Effective rate of the 2-stage cascaded dissipation process, evaluated from the smallest imaginary component of the eigenvalues of matrix Eq.~(\ref{eq:diss_matrix}). %, which is a good proxy for the rate of the 2-stage cascaded dissipation process. 
        The $\times$ denotes the drive rates used in the experiment. (\textbf{C}) The transmon $|e\rangle, |f\rangle$ population for the three different cavity initial states as in \textbf{A} over time, using the same master equation fit as \textbf{A}. (\textbf{D}) Wigner tomography for prepared initial even-parity superposition states (top) and the odd-parity states after 15 $\mu$s of PReSPA drives (bottom). There is a deterministic phase-space rotation due to the self-Kerr of the cavity.
        %with a phase accumulated from the self-Kerr rotation. 
        %The off diagonal elements of the reconstructed density matrix describe the coherence of the process. 
        The coherence preservation factor, defined by the ratio of the corresponding off-diagonal density matrix elements, is stated next to the arrows between the Wigner functions. All initial states are prepared using numerically-optimized control pulses~\cite{heeres2017implementing}. 
        }
	\label{fig:fig2} % give each figure a logical label name
\end{figure*}

\begin{figure*}[tbp] % Do NOT use \begin{figure*}
	\centering
	\includegraphics[width=5.0in]{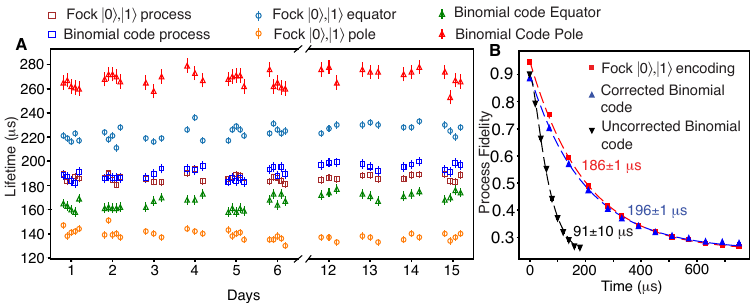} 

	\caption{\textbf{Logical qubit lifetime under AQEC.}
		(\textbf{A}) The state and process fidelity lifetimes for the corrected binomial code using PReSPA and the $|0\rangle, |1\rangle$-Fock encoding, as measured over multiple days. Pole and equator state fidelities are fit to an exponential decay lifetime with error bars determined from fit uncertainties. The process fidelity lifetimes for both the encodings are computed as the inverse of its decay slope at at $t\rightarrow0$, or $\tau_{process} = 1/\left(\frac23 T_{eq}^{-1} + \frac13 T_{p}^{-1}\right)$. (\textbf{B}) Process fidelity over time of the binomial code with and without correction along with the $|0\rangle, |1\rangle$-Fock encoding, averaged over all data acquired in the last four days of measurement in \textbf{A}. The dashed red and blue curves are sum of two exponential functions two time scales $T_{eq}$ and $T_p$, computed from independent exponential fits of the state fidelity of pole and equator states. The uncorrected binomial code is fit to a sum of two exponential functions, taking into account partial information recovery from the decoding process when the cavity has lost only one photon, with the $1/e$ decay time shown in the plot.}
	\label{fig:fig3} 
\end{figure*}

\section*{Experimental Results}

To experimentally tune up the optimal drive rates $\Omega_1$, $\Omega_2$ and to characterize the effectiveness of the dissipative photon addition process, %achieved in the experiment for different amplitudes, 
we measure the time-domain dynamics of the storage photon numbers under the combined drives, starting in different initial even-parity Fock states (Fig.~\ref{fig:fig2}A). The photon number dependent dispersive shift of the transmon allows for tracking of the $\ket{n}$-state probability of the storage cavity.  This is done by measuring the transmon after exciting its $|g\rangle\rightarrow|e\rangle$ transition with a spectrally selective microwave pulse centered at frequency $\omega_q-n\chi_{ge}$. The population data over time is fit to the numerical solution of the master equation with Hamiltonian $\hat{H}_d$ and dissipators $\mathcal{D}[\hat{r}], \mathcal{D}[\hat{q}], \mathcal{D}[\hat{a}]$ (with corresponding jump rates), for each even-to-odd parity Fock state conversion, with $\Omega_1$ and $\Omega_2$ as fit parameters.  In this scheme, the transmon is partially excited to the $|f\rangle$-state that might decay to $|e\rangle$ during the correction cycle, which is also tracked and included in the fit (Fig.~\ref{fig:fig2}C).  The combs of FWM drives induce an AC Stark shift in the transmon frequency, which is characterized in separate spectroscopy measurements and taken into account in the frequencies of the applied PReSPA drives. % in an iterative process to optimize the experimental drive amplitudes for the fastest dissipation process.

The ability to achieve relatively fast continuous QEC rates at low drive power is a notable technical advance of this AQEC demonstration.  
From Fig.~\ref{fig:fig2}A, we achieve a dissipation halftime $\tau_{cor}\approx 4$ $\mu$s, which is sufficiently fast in comparison to the optimal active QEC cycle time in most previous studies of bosonic QEC~\cite{hu2019quantum, ofek2016extending, sivak2023real}. This PReSPA operating condition fits to drive rates of $\Omega_1/2\pi\approx55$ kHz and $\Omega_2/2\pi\approx160$ kHz for each of the three conversion paths, while the total ac Stark shift of the transmon (which is often quoted as a proxy for parametric drive strengths) is less than 100 kHz for all drives combined.  This Stark shift is about two orders of magnitude lower than the onset of prominent spurious transitions revealed in recent studies~\cite{dai2025spectdust, xia2025exceedingparametric},
%mori2025suppress}, 
and places our driven system deep in the lowest-order perturbative regime of Josephson non-linearity. 
Indeed, we observe no multi-tone mixing effects, the required amplitude of each drive tone to yield the same FWM Rabi rates simply scales as the bosonic factor, % without the need for individual tuning, 
and each FWM tone contributes an additional Stark shift of $\Delta_{ss1}=\Omega_1^2/n\chi_{ge}$ and $\Delta_{ss2}=\Omega_2^2/\chi_{qr}$ as predicted by perturbation theory (see Supplementary Text).  Minimal  
Stark shift also makes the experiment insensitive to temporal drifts of microwave power and eliminates the need for frequent calibration.  Nevertheless, we still observe a very small transmon heating effect, which we will revisit in the Discussion section.  Due to the unfortunately low cavity $T_1$ time of 136 $\mu$s in this experiment, the conversion fidelity of the photon addition decreases substantially from $n=$ 1 to 3 and 5, which limits the overall performance of this AQEC demonstration.

In another set of measurements, we apply PReSPA drives on a pairwise superposition of even-parity Fock states to show preservation of coherence after the photon addition. We perform direct Wigner function tomography using the transmon ancilla to characterize both the initial state and the converted state after applying the drives for 15 $\mu$s (Fig.~\ref{fig:fig2}D) for comparison. Density matrix reconstruction informs that PReSPA produces the corresponding odd-parity states with well-preserved coherence. As the mean photon number increases for the initial even-parity superposition states, the preserved coherence decreases primarily from competing photon loss events.  %is not as well preserved in the pumped odd-parity superposition state.

We use the process fidelity to quantify the overlap between the experimentally realized quantum map ($\chi_{exp}$) and the identity map ($\mathcal{I}$) on the logical qubit, thereby providing a comprehensive measure of how accurately PReSPA preserves the encoded information over time. %in the dissipative process. 
The process fidelity, formally defined as $\mathcal{F}_{process}=$ Tr$[\chi_{exp}\mathcal{I}]$, is linearly related to the experimentally measured average state fidelity $\mathcal{F}_{avg}$ of the six cardinal states of the logical Bloch sphere by $\mathcal{F}_{process}=0.25 + 1.5(\mathcal{F}_{avg}-0.5)$. Because of the difference in exponential decay times of the pole states $T_p$ and equator states $T_{eq}$, the process fidelity decays as a sum of two exponential functions. %However, at $t\ll\tau_{process}$ the fidelity time dynamics are approximated to be a single exponential and decay rate becomes a weighted average, 
Since the decay rate of the process fidelity at the short-time limit is the most relevant metric for the QEC performance, we will quote its inverse as the logical lifetime $\tau_{process} = 1/\left(\frac23 T_{eq}^{-1} + \frac13 T_{p}^{-1}\right)$.

%To confirm the reproducibility and stability of PReSPA, 
We tracked the fidelity lifetimes ($T_p, T_{eq}, \tau_{process}$) of the corrected binomial code and the best physical qubit of system, a Fock $\{|0\rangle, |1\rangle\}$ encoding, for multiple days, as shown in Fig.~\ref{fig:fig3}A.  The AQEC remains effective with no recalibration of PReSPA drives throughout the experiment. %except for adjusting for some slow fluctuations of the bare transmon frequencies.  
Averaged over the whole data set, the binomial code shows a mean process fidelity lifetime of $\tau_{process}=191\pm1\mu$s (standard error), which exceeds the Fock $|0\rangle, |1\rangle$-encoding,  $\tau_{process}^{|0\rangle|1\rangle}=186\pm1\mu$s, by 3\%. 
Close inspection of Fig.~\ref{fig:fig3}A reveals small but statistically-significant changes of both the raw physical coherence properties and the logical qubit lifetime over days. Notably, over the last four days, the binomial code under AQEC showed a mean lifetime of $\tau_{process}=196\pm1$ $\mu$s, about 5\% beyond its breakeven reference. This improvement may be attributed to a slightly lower excited-state population and better $T_1$ time of the transmon.

To further illustrate the performance of the continuous AQEC, we plot in Fig.~\ref{fig:fig3}B the time decay of the process fidelity of the corrected binomial code against the break-even reference and the uncorrected binomial code, for data averaged over the last four days. Compared to the binomial encoding under free evolution (black), the AQEC provides an improvement of about 2.2 times. The infidelity of the state preparation and decoding leads to lower initial process fidelity for the corrected binomial code (blue) when compared to the Fock encoding (red).  The two decay curves cross at $t\sim400$ $\mu$s, showing that we can indeed recover slightly more information from the protected logical qubit than from the physical qubit at long times after accounting for the overhead of encoding and decoding.

\begin{figure} % Do NOT use \begin{figure*}
	\centering
	\includegraphics[width=8.5cm]{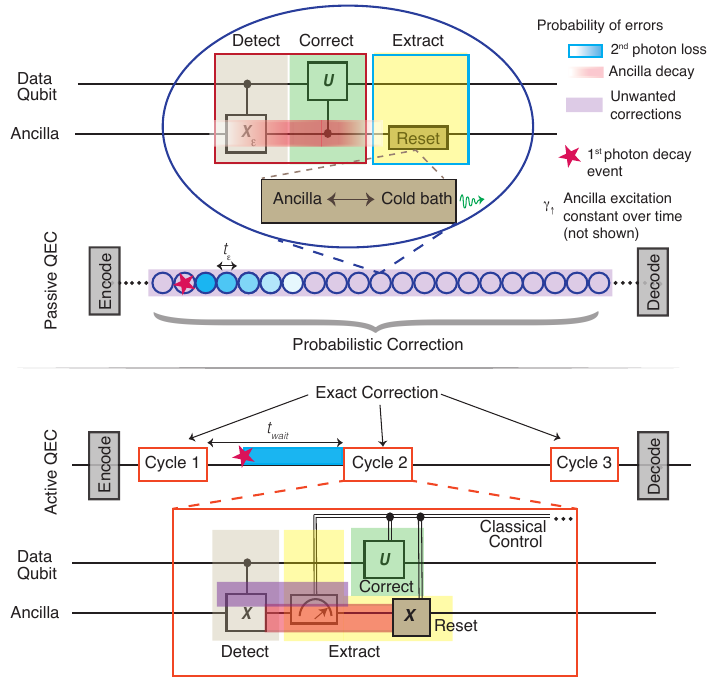} 
	\caption{\textbf{Comparison of the actions and residual error channels in continuous AQEC vs active QEC.}
        Illustration showing an overview of the passive (top) and active (bottom) QEC schemes. The small circles in the passive case show the correction happening continuously after the error occurs, where each circle can be viewed as a finite probability in the small time step ($t_{\epsilon}$) of the correction process, outlined in the circuit above, occurring. The discrete error correction cycles in the active case are shown, outlined by the orange rectangles. A zoomed in circuit diagram for these measurement based correction cycles is shown below. The uncorrectable error pathways are shown as shaded regions, with specific errors outlined in legend in the upper right corner, where the more opaque regions correspond to larger instantaneous error rates. For example, after an initial error, the probability of a second error decreases over time in passive case thanks to the increasing chance of having completed the correction, %as time progresses in the passive case, the chance that the correction process occurred increases, thus lowering the probability of a second photon loss before the first loss is corrected, 
        shown as a progressively lighter blue shade over time. In the active case, the probability is a constant until reaching the next active QEC operation, thus shown by the constant blue shading, .  %between correction cycles due to the fixed times that the correction occurs.
        }

	\label{fig:fig4} % give each figure a logical label name
\end{figure}

\section*{Discussions}

The continuous AQEC demonstrated here is distinct from the effect of driven dynamical protection.  The latter is responsible for protecting spin-locked or Floquet qubits~\cite{huang2021engineering, yan2013rotating,
nguyen2024programmable} and plays a major role in dissipative-cat or Kerr-cat qubits~\cite{hajr2024high, grimm2020stabilization, reglade2024quantum, frattini2024observation}.  Dynamical protection exploits the (typically low-frequency) spectral feature of the noise, using an engineered effective energy gap or the quantum Zeno effect to prevent errors that cumulate from noisy continuous quantum dynamics over time.  The noise spectrum of cavity photon decay, on the other hand, is mostly white over the practically-relevant spectral range. %(except for small fluctuations caused by two level systems)
Its dynamics is better understood as comprised of intrinsically discrete quantum jumps as a result of the Fermi's golden rule coupling to a high density of environmental states, which is unaffected by the PReSPA drives. 

Despite the continuous nature of the AQEC, its underlying steps of actions have a direct correspondence with the active QEC paradigm based on measurement cycles, as illustrated in Fig.~\ref{fig:fig4}.  Continuous AQEC acts as repeated applications of probabilistic correction cycles over infinitesimal times steps~\cite{ahn2002continuous, AhnDoherty2013book}.  Similar to active QEC, each cycle is composed of conditional excitation of the ancilla for syndrome detection, a correction unitary conditioned on the ancilla state, and an ancilla reset to extract the entropy.  The probabilistic nature of the correction lies in an effective weak measurement of the syndrome via the ancilla, as opposed to the strong measurement in active QEC cycles.  Furthermore, the AQEC applies an ancilla-controlled unitary on the encoded qubit (controlled-$U$) for the recovery operation, as opposed a direct $U$ activated by classical logic after measurement.  It is worth noting that a hybrid approach of the two paradigms have employed discrete cycles of unitary correction and dissipative reset without measurement-feedback%, as recently demonstrated in Refs.
~\cite{lachance2024autonomous, de2022error, sun2025extending, ni2025autonomous, li2025beating, debry2025errorcorrectionlogicalqubit}. 

Figure~\ref{fig:fig4} also facilitates an intuitive comparison of main sources of imperfections in our passive AQEC protocol versus active QEC.  The susceptibility to uncorrectable $2^{nd}$-order photon-loss errors depends on the average time the system spends in the error space after a first photon loss, which is $\tau_{cor}$ and $\tau_{wait}/2$ respectively.  Speeding up the correction % i.e.~reducing these characteristic correction times, 
incurs a tradeoff of increased probabilities of injecting uncorrectable errors.  A main type of injected errors can be understood as spurious ``unwarranted corrections'' that drives the code states into error states, which in the case of AQEC is caused by reduced parity-subspace selectivity under stronger dissipation and in the case of active QEC is caused by increased gate and readout errors under more frequent QEC cycles.  Assuming %the use of a static dispersively-coupled ancilla for syndrome extraction, %(as in existing demonstrations of bosonic QEC),
a static dispersive coupling between the data and ancilla qubits,
both protocols are susceptible to forward propagation of ancilla decay error during correction. %cycle until the ancilla rest.  
Similarly, both types of QEC are susceptible to any spontaneous excitation of the ancilla at all times.  

In the Supplementary Information, we present a detailed analysis of these remaining error channels in our binomial qubit in comparison with its closest actively-corrected counterpart as studied in Ref.~\cite{ni2023beating}.  Assuming the transmon readout performance therein and equal level of hardware coherence, we find that the two protocols provide overall similar performances in corrected logical qubit lifetimes with slight differences in error-budget tradeoffs. 
The main performance advantages of AQEC are lower impact from unwanted corrections from parity-measurement errors and recovery unitary control errors. %, which benefits from the elimination of ancilla readout error and a suppression of parity-mapping gate error. 
The main advantages of active QEC are more flexibility to reduce transmon decay errors and the ability to correct for the distortion of the bosonic codeword coefficients. %, which benefits from the decoupling of its characteristic correction time $\tau_{wait}$ and the ancilla reset time (several multiples of $1/\kappa$).  
Given the hardware coherence in the present device, AQEC slightly outperforms active QEC in our analysis, which shows the competitiveness of this passive strategy in addition to its hardware efficiency in eliminating the technically demanding high-fidelity readout and fast feedback.

Moving forward, the appealing convenience of the passive QEC strategy for superconducting circuit QED comes at the cost of managing always-on parametric drives in the background. % and its potential negative impacts.  
It is well-known that strong off-resonance drives in Josephson circuits can lead to drive-induced unwanted state transitions (DUST)~\cite{dumas2024mist, dai2025spectdust, xia2025exceedingparametric}, which limits fast readout and parametric gate operations~\cite{xia2025exceedingparametric} and is also a major bottleneck for engineering driven qubits~\cite{frattini2024observation, nguyen2024programmable}. Despite working in the weak-drive regime, we observe a slow increase of the excited-state occupancy of the transmon over long timescales, translating to an excess transmon heating rate over equilibrium of $\Delta\gamma_\uparrow \approx 0.4$ ms$^{-1}$ under our AQEC conditions.  Its impact on our reported logical qubit lifetime is insignificant, but will become a major trade-off factor in devices with much better cavity $T_1$ times.  In Supplementary text, we present a systematic study of this millisecond-scale DUST effect at low drive power, which shows an unexpected near-linear dependence on drive amplitude and no apparent memory effect (which should exclude extrinsic factors such as heating of off-chip filters or attenuators).  The mechanism of the heating effect remains elusive and is an important topic of future study in order to advance continuous AQEC techniques for cavity bosonic qubits.  Future directions further include phase-space stabilization by engineering four-photon dissipation~\cite{vanselow2025dissipatingquartets, mirrahimi2014dynamically} and incorporation of logical gate operations~\cite{ma2020error, reinhold2020ecgates}.

In conclusion, we have experimentally demonstrated a passively and continuously corrected quantum memory with logical lifetimes reaching breakeven. Although there is no obvious thermodynamic description of this driven system as in the true sense of self-correction~\cite{brown2016qmtemmp}, we note that our logical qubit is protected under a time-independent Hamiltonian and dissipation processes in a pre-defined rotating frame.  Our demonstration is, to the best of our knowledge, the first example that quantum information stored in the collective state of elementary excitations under a quasi-equilibrium environment persists beyond the limit imposed by the excitation lifetime.

%\clearpage 

%%%%%%%%%%%%%%%% ACKNOWLEDGEMENTS %%%%%%%%%%%%%%% 

%\subsection*{Acknowledgments}
\textit{Acknowledgments} -- 
We thank J.~M.~Gertler for contributions to the device design, B.-J.~Liu and Y.-Y.~Wang for helpful discussions. %, and \cw{xxx} for experimental assistance.  
This research was supported by the US Department of Energy (DE-SC0021099). We also acknowledge partial support from Defense Advanced Research Projects Agency (HR00112490360).

%%%%%%%%%%%%%%%% REFERENCES %%%%%%%%%%%%%%%

\bibliography{science_template}

%%%%%%%%%%%%%%%% END OF MAIN TEXT %%%%%%%%%%%%%%%

\newpage

%%%%%%%%%%%%%%%% START OF SUPPLEMENT %%%%%%%%%%%%%%%

\renewcommand{\thefigure}{S\arabic{figure}}
\renewcommand{\thetable}{S\arabic{table}}
\renewcommand{\theequation}{S\arabic{equation}}
\renewcommand{\thepage}{S\arabic{page}}
\setcounter{figure}{0}
\setcounter{table}{0}
\setcounter{equation}{0}
\setcounter{section}{1}
\setcounter{page}{1} % not 0 as \newpage already started a supplementary page
% References continue the numbering from the main text.

%%%%%%%%%%%%%%%% SUPPLEMENT TITLE PAGE %%%%%%%%%%%%%%%

\begin{titlepage}
    \begin{center}
        {\Large \textbf{Supplementary Information}}\\[0.5em]
        
        {\large \textbf{Passive quantum error correction of Photon Loss at Break-even}}\\[1em]
        {Shruti Shirol, Sean van Geldern, Hanzhe Xi, Chen Wang\\
        Department of Physics, University of Massachusetts-Amherst, Amherst, MA 01003}
        \rule{\textwidth}{0.4pt}\\[1em]
    \end{center}

\end{titlepage}

\tableofcontents

\newpage
%%%%%%%%%%%%%%%% MATERIALS AND METHODS %%%%%%%%%%%%%%%

\subsection{Materials and Methods}

\subsubsection{Qubit and package setup}
We use a 3D-planar hybrid circuit QED architecture as introduced in \cite{axline2016architecture}. The 3D package is made with high purity (99.9995\% pure) Aluminum that undergoes a chemical etching to smooth out surface irregularities, thus increasing the internal Q for modes within. The package is built to have a vertical cylindrical waveguide with a post at its center to make a 3D high-Q $\lambda/4$-resonantor mode in the system Fig.~\ref{fig:fig2}, and a horizontal cylindrical tunnel where a sapphire chip can be inserted. A fixed-frequency transmon qubit is fabricated on a double-side-polished sapphire chip using the Dolan bridge method. With double-angle evaporation in plassys we realize a Josephson junction Al/AlO$_x$/Al that has the first layer of aluminum at 40nm thickness and the second one at 60nm. The oxide layer is about 1-2 nm resulting from 10 minutes of oxidation at 100 mbar. On the same chip, we also make a 2D (low-Q) stripline resonator that is used for readout and as a cold bath in the AQEC. The corners of the metal strip are rounded to reduce the surface dielectric loss. The transmon capacitively couples with a high-Q, ($\chi_{ge}/2\pi = 1.15$ MHz, $\kappa_a/2\pi=1.1$ kHz) and low-Q ($\chi_{qr}/2\pi = 1.13$ MHz, $\kappa_r/2\pi=0.58$ MHz) resonator modes. The coupling between the storage and reservoir mode is undesired and $\chi_{ar}/2\pi\approx$ 9.3 kHz. The chip is held in place using a clamp that contains deformable indium. In order to increase (decrease) the $\chi_{ge}$ we add (reduce) more indium in the clamp that pushes (pulls) the chip towards (away) the post cavity.   

\subsubsection{Fridge and Measurement setup}
The package is mounted on the bracket inside an amuneal can on the mixing chamber. The fridge setup is shown in supplementary figure \ref{fig:fridge_diagram}. With this configuration, we achieved the qubit and cavity coherence values stated in Table \ref{tab:system_params}. We have seen an improvement in the transmon relaxation and spurious excitation rates by adding eccosorb filters, a 30 dB attenuator on the mixing chamber, and an eccosorb foam on the walls of the amuneal can. The eccosorb filters improve qubit $T_{1q}$ by lowering the incident IR radiation \cite{connolly2024eccosorb}. The room temperature microwave control setup we have is typical for cQED architectures where the transmon and cavity drives are generated using sideband mixing of a local oscillator tone (LO) and intermediate frequency tone (IF) from an arbitrary waveform generator (AWG). We transmit the readout drive from transmon input port to readout output port to probe the transmon state. The transmon input port is also used to drive the transmon. The readout signal exiting the device is amplified at 4K with a HEMT that has 40 dB gain, followed by a second amplifier with 35 dB gain at room temperature. The signal is demodulated to 50 MHz and amplified again before going to the analog-to-digital converter (ADC). The readout fidelity is 88\%, which would be insufficient for high-fidelity single-shot readout but suffices for the AQEC experiment. Future experiments could improve on readout fidelity with a parametric amplifier and incorporate the mid-circuit measurements for erasure detection. The difference in the measured voltages for the transmon in state $|g\rangle$ versus $|e\rangle$ is the raw contrast and is used to scale all the measurements thereafter to determine the cavity and transmon excitation probabilities. Since we have a non-zero excited state population (1.7\%) of transmon in the equilibrium, we appropriately scale the raw contrast value to approximately equal full $|g\rangle-|e\rangle$ contrast. Details of how the measurements in Fig.~\ref{fig:fig2}A are scaled are discussed in the following sections.

\subsubsection{Parameter choices}
%kappa_r is not mentioned
The frequency placements and coupling strengths are shown in the parameter list presented in Table \ref{tab:system_params}. The transmon, storage cavity, and readout mode frequencies are chosen so that the required four-wave mixing combs are in a suitable range for the microwave setup. The first and second stages of AQEC use pumping tones at frequencies $\omega_{d1}=\omega_a+2\omega_{ge}-\alpha -n\chi_{gf} = 11.49$ GHz and  $\omega_{d2} = \omega_r - (2\omega_{ge}-\alpha -n\chi_{gf} ) = 1.89$ GHz, respectively for the mode frequencies as listed in Table \ref{tab:system_params}. We place the mode frequencies such that the FWM frequencies required for PReSPA do not overlap with other modes in the system or induce multiple-photon processes. Since PReSPA drives the transmon between $|g\rangle $ and $|f\rangle$ state, which provides intrinsically stronger interaction due to higher participation of the non-linear transmon mode. We can drive at low enough strengths that produce Stark shifts $\ll\Omega_1,\Omega_2$. For these weak drives we do not expect multi-photon transitions. The drives also produce consistent Rabi rates without the need for frequent tune-up procedures since the temporal drifts in amplitude or frequency of the drives do not affect the experiment conditions. There is also no parametric mixing effects from multiple tone (e.g., as in \cite{gertler2021protecting}). 

The PReSPA drive rates for first and second stage AQEC follow the hierarchy $\Omega_1<\Omega_2\ll\chi$ and thus selectively drive when the logical qubit is in the error space manifold. A stronger storage-transmon dispersive coupling strength allows faster correction rates since it can accommodate larger $\Omega_1$. However, increasing $\chi_{ge}$ also increases the other higher-order terms of the Hamiltonian such as self-Kerr $K$ and $\chi_{q}'$ which contribute to error budget of the correction scheme. By using the state $|f\rangle$ of the transmon in the PReSPA,  since now $\Omega_1\ll\chi_{gf}$ where $\chi_{gf}=\chi_{ge}+\chi_{ef}$, a larger $\Omega_1$ can be accommodated without increasing the coupling between storage and transmon. With the reservoir loss rate $\kappa_r/2\pi=0.58$ MHz, a critically damped driving is observed at $\Omega_1=55$ kHz $\Omega_2=160$ kHz. A sweep of pumping rates to find optimal dissipation on the error state is shown in Fig.~\ref{fig:fig2}B, and is already discussed in the main text. 

\subsubsection{Four wave mixing tones generation and calibration}

The two four-wave mixing combs are sideband modulated using two separate Local Oscillators (LO) and Intermediate Frequency (IF) channels from an AWG for IQ mixing. The three square pulses with 100 ns ramp up and ramp down time with frequency $\chi_{gf}, 3\chi_{gf}, 5\chi_{gf}$ are digitally generated from the AWG IQ pair and mixed with the LO drives to produce $\omega_{d1}-\chi_{gf}, \omega_{d1}-3\chi_{gf},\omega_{d1}-5\chi_{gf}$ simultaneously. The IF drives from another pair of AWG channels mix with the second LO drive to produce $\omega_{d2}+\chi_{gf}, \omega_{d2}+3\chi_{gf}, \omega_{d2}+5\chi_{gf}$ simultaneously. These signals are combined using power dividers, amplified, and driven through the qubit input port. 

To find the transmon Stark shift in the presence of the AQEC drives, we perform a spectroscopy sweeping the drive frequencies such that $\omega_{d1}+\omega_{d2}$ is constant and measure the transmon conditional on $|1\rangle$-state cavity population. The applied drives have frequency $\omega_{d1}+\Delta f$ and $\omega_{d2} - \Delta f$ with varying $\Delta f$. The Stark shift is predicted from finding this detuning $\Delta f$ at which the $\ket{1}$ state population maximizes. As discussed in the main text, the tones in the first comb are scaled by $1/\sqrt{n}$ (where $n=1,3,5$) to compensate for the bosonic factors. The second comb has constant amplitude for all the tones across photon numbers since there is no bosonic enhancement factor to compensate for.  

\begin{table}[t] % Do NOT use \begin{table*}
	\centering
	% Captions go above tables
	\caption{\textbf{System Parameters}
		This table lists all the mode frequencies, coupling strengths between modes, and coherence properties of each mode. We only state the most relevant and significant fourth-order coupling terms of the Hamiltonian. The higher-order (sixth-order and beyond) terms are negligible except for $\chi_{q}'$ which becomes important at higher storage photon populations and thus added to the Hamiltonian for numerical pulse optimization used for initial state preparation.}
        
	\label{tab:system_params} % give each table a logical label name
	
	\begin{tabular}{lccc} % four columns, alignment for each
		\\
		\hline
		Parameter & Symbol & Value\\
		
		\hline
		Transmon frequency  & $\omega_{q}/2\pi$ & 3482.9 MHz \\
		Storage Cavity frequency  & $\omega_{a}/2\pi $ & 4657.9 MHz \\
            Readout frequency  & $\omega_r/2\pi$ & 8725 MHz\\
            \hline
            Transmon Anharmonicity  & $\alpha/2\pi$ & 134.28 MHz\\
            Storage Cavity anharmonicity  & $K/2\pi$ & 3.3 kHz\\
            Storage-Transmon $g-e$ coupling  & $\chi_{ge}/2\pi$ & 1.12 MHz\\
            Storage-Transmon $e-f$ coupling  & $\chi_{ef}/2\pi$ & 0.95 MHz\\
            Storage-Transmon $6^{th}$-order coupling  & $\chi_{q}'/2\pi$ & 1.9 kHz\\
            
            Readout-Transmon coupling  & $\chi_r/2\pi$ & 1.13 MHz\\

            Readout decay  & $\kappa_r/2\pi$ & 0.58 MHz\\
            \hline
            Transmon $|e\rangle \rightarrow |g\rangle$ decay  & $T_{1q}$ & 50 $\mu$s \\
            Transmon $|f\rangle \rightarrow |e\rangle$ decay & $T_{1ef}$ & 31 $\mu$s \\
            Transmon $T_{2}$ Ramsey  & $T_{2R}^{*}$ & 53 $\mu$s \\
            Transmon $T_{2}$ Echo & $T_{2E}$ & 70 $\mu$s \\
            Transmon  $g-f$ $T_{2}$ Ramsey & $T_{2gf}^{*}$ & 30 $\mu$s \\
            Transmon $|e\rangle$ population &  & 1.7\% \\  
            \hline
            Storage Cavity $|1\rangle \rightarrow |0\rangle$  & $T_{1a}$ & 136 $\mu$s \\
            Storage Cavity $T_2$ & $T_{2a}$ & 235 $\mu$s \\ 
            Storage Cavity $|1\rangle-$state population  &  & 0.6\% \\ 
		\hline
        
	\end{tabular}
\end{table}

\begin{table*}[t] % Do not use \begin{table*}
	\centering
	% Captions go above tables
	\caption{\textbf{Parameters of the two FWM drive transitions}
		This table shows the measured Stark shifts ($\Delta_{ss}$) and drive strengths ($\Omega_{n}$ meas.) along with the predicted $\beta_{n}$ and drive strengths ($\Omega_{n}$ est.), inferred from the Stark shift value, for the two stages of FWM tones. Note that these values are all only for the tones used in the $\ket{0}\rightarrow \ket{1}$ storage addition process.}
	\label{tab:drive_params} % give each table a logical label name

	\begin{tabular}{lcccc} % four columns, alignment for each
		\\
		\hline
		FWM drive & $\Delta_{ss}$(kHz) & $\beta_{ n}$ & $\Omega_{n}$ est. (kHz)& $\Omega_{n}$ meas. (kHz)\\
		\hline
		$|0, g, 0\rangle\rightarrow|1, f, 0\rangle$ & 3 & 0.0033 & 57 & 55 \\
        $|1, f, 0\rangle\rightarrow|1, g, 1\rangle$ & 32 & 0.011 & 190 & 160 \\
        \hline

		\hline
	\end{tabular}
\end{table*}

\subsubsection{State Preparation}

For various PReSPA calibration and characterization experiments (e.g., Fig.~\ref{fig:fig2}A, C, D), we prepare even-parity Fock states and a superposition of even-parity Fock states. An Optimal Control Theory (OCT) pulse, optimized using a gradient-based optimizer from Q-CTRL \cite{boulder_opal}, is used to generate these arbitrary states in the storage cavity. We also used OCT pulses to prepare the six Bloch sphere cardinal states and a decoding pulse that maps the cavity state to the transmon state which is then characterized to get logical state fidelity and eventually the PReSPA process fidelity in Fig.~\ref{fig:fig3}(A, B).

\subsubsection{Cavity and transmon population measurement}

We probe the photon number in our storage cavity using the transmon ancilla. We use a Gaussian pulse 2.4 $\mu$s long ($4\sigma$) that is selective in frequency, allowing us to drive the ancilla conditioned on different Fock states in the storage cavity. In order to get the data in Fig~\ref{fig:fig2}A we read the transmon state after PReSPA pumping in two distinct measurements. In the first one, we do not use any transmon drive pulse and read the background right after applying the pump tone. In the consecutive measurement, we use a selective Rabi $\pi_{ge}$ pulse centered at $\omega_q-n\chi_{ge}$ after the PReSPA pumps that excites the transmon to $\ket{g}\leftrightarrow\ket{e}$ conditional on population in cavity. We can represent the readout values from these experiments in terms of the equations $P_{\ket{n,g}}A+P_{\ket{n,e}}B+P_{\ket{n,f}}C=Background$ (readout value without transmon pulse) and $P_{\ket{n,e}}A+P_{\ket{n,g}}B+P_{\ket{n,f}}C=Data$ (readout value with transmon pulse), where $P_{\ket{n,g}}, P_{\ket{n,e}}, P_{\ket{n,f}}$ are the populations of the respective states and $A, B, C$ are the raw readout values measured when the transmon is initialized in the state $\ket{g}, \ket{e}, \ket{f}$. Subtracting the two equations and dividing by $A-B$ gives the plotted data, $P_{\ket{n,g}}-P_{\ket{n,e}}=(Background-Data)/(A-B)$. 

This measurement is repeated for all the even-parity Fock initial states and is presented in Fig.~\ref{fig:fig2}A. The population difference  $P_{\ket{n, g}}-P_{\ket{n, e}}$ from the numerical solution of the master equation with Hamiltonain $\hat{H_d}$ (Eq.~\ref{eq:ham_drive}) and dissipators $\mathcal{D}[\hat{r}],\mathcal{D}[\hat{a}], \mathcal{D}[\hat{q}] $ is used to fit this data for all $n=1,3,5$ to extract the drive rates $\Omega_1, \Omega_2$. The fit yields the populations after $t=15$ $\mu$s of PReSPA drives (near convergence) to be $P_{\ket{1,g}} = 0.92, P_{\ket{1,e}} = 0.04$ for the initial state of $\ket{0, g}$. % with the dashed-line fit in Fig.~\ref{fig:fig2}A representing $P_{\ket{1,g}}-P_{\ket{1,e}}$.  
The populations for other odd-parity Fock states after $t=15$ $\mu$s of correction drives for respective even-parity Fock states are $P_{\ket{3,g}} = 0.78, P_{\ket{3,e}} = 0.04$, and $P_{\ket{5,g}} = 0.67, P_{\ket{5,e}} = 0.045$.  These $\ket{e}$-state populations result from a transmon $\ket{f}\rightarrow\ket{e}$ decay in the intermediary step of PReSPA, which terminates the correction prematurely but nonetheless accomplishes a photon-addition process but with a random phase.  %However, since the photon has already been added at this stage, 
Based on this analysis, the data in Fig.~\ref{fig:fig2}A may have underestimated the success probability of PReSPA ($\sim P_{\ket{n,g}}$) by up to 4\% and the actual cavity photon population $P_{\ket{n}}$ by up to 8\%.  We presented the conservative estimates since we did not measure $P_{\ket{n,e}}$ at the time (which is possible by using number-selective $\ket{e}-\ket{f}$ pulses). 

We also measure the intermediary transmon states using a protocol similar to transmon thermal population measurement, which assumes higher states are not occupied~\cite{jin2015thermalresidue, geerlings2013drivenreset}. A sequence of measurements with Rabi $\pi$-pulses driving $\ket{f}-\ket{h}$ and $\ket{e}-\ket{h}$ measures the population in $\ket{f}$ and $\ket{e}$.

\subsubsection{Measuring the fidelity of logical state}

The performance of PReSPA on the logical codewords is gauged by measuring the fidelity of logical states against the ideal Bloch sphere cardinal states, which can then be used to compute the process fidelity of the correction map. As discussed in the main text, the lifetime $\tau_{process}$ is calculated from the pole ($T_p$) and equator ($T_{eq}$) logical state lifetimes. When ($t\ll T_{1a}$), the process fidelity can be estimated to have a single exponential decay, with a rate that is a weighted average of pole and equator logical state decay rates. The inverse of this average gives the lifetime, $\tau_{process} = \frac{6}{4/T_{eq} + 2/T_{p}}$.

To determine the fidelity of each Bloch sphere cardinal state, the logical state is decoded to the ancilla, which can be characterized with tomography. We use a decoding OCT pulse that performs the following map,  
\begin{equation}
    |u_{0}, g\rangle \rightarrow |0, g\rangle,
    |v_{0},g\rangle \rightarrow|0, e\rangle,
    |u_{1}, g\rangle \rightarrow |1, g\rangle
    \label{eq:ham_dec2}
\end{equation}
where $u_0=\frac{\ket{1}+\ket{5}}{\sqrt{2}}$, $u_1=\frac{\ket{1}-\ket{5}}{\sqrt{2}}$	and $v_0=\ket{3}$. The decoding was also designed to recover states in the error space after single-photon losses, even though it was not as perfect as for the code-space states.

The pole state fidelity is a straightforward measure of the transmon population but equator states require a frame change where the superposition of $\ket{0_L}$ and $\ket{1_L}$ are stationary. Using Ramsey-like measurement,we find the rotation frequency of $\ket{0_L}$ and that of $\ket{1_L}$ with respect to the state $\ket{0_L}$. A correct phase on OCT pulse used for decoding changes the frame to that where $\ket{0_L}$ is stationary. After decoding, in the transmon state tomography step we drive the transmon around axis adjusted by the rotation angle between $\ket{0_L}$ and $\ket{1_L}$ at the time of decoding,  which is now the angle between $\ket{g}-\ket{e}$. This is repeated for all the four equator states which are then averaged to quote the equator lifetime $T_{eq}$

%\section*{Supplementary Text}

\subsection{Model and optimization of two-stage cascaded dissipation}
For efficient pumping, we choose the drive rates that satisfy the critical damping condition for a two-step dissipation scheme, yielding the fastest non-oscillatory relaxation of the logical state back to the code-space. This condition can be determined by studying a simplified version of the full master equation governing the system, which is effectively a non-Hermitian Hamiltonian~\cite{reiter2012effective}. We can treat $\ket{n-1,g,0},\ket{n,f,0},\ket{n,g,1}$ as a manifold of excited states and $\ket{n,g,0}$ as ground state for all $n=1,3, 5$. 
Since there are no interactions in the ground state manifold to consider when optimizing the correction process, we can write an effective non-Hermitian Hamiltonian ($\hat{H}_\text{eff}$) as, 
\begin{equation}
\hat{H}_\text{eff}=\hat{H}_{d}-\sum_{n=1,3,5}\frac{i\kappa_{r}}{2}\ket{n,g,0}\bra{n,g,1}
\label{eq:Heff}
\end{equation} 
The $\Omega_1, \Omega_2$ are required to match for all the $n$, thus the dynamics for all transition paths are identical. We may use a single $3\times3$ non-Hermitian Hamiltonian matrix to describe each of the three dissipative processes:

\begin{equation}
\hat{H}_\text{diss}=
\begin{pmatrix}
0 & \Omega_1 & 0\\
\Omega_1^* & 0 & \Omega_2\\
0 & \Omega_2^* & -i\kappa_r/2\\
\end{pmatrix}\label{eq:diss_matrix}
\end{equation} 
 
The imaginary component of the eigenvalue of the eigenstate that has the largest overlap with $\ket{n-1,g,0}$, approximately gives the effective dissipation of error state back to code state. In the absence of the $\Omega_{1}$ drive, this reduces to the familiar three level $\lambda$ system, which has a critical damping at $\Omega_{2}=\kappa_{r}/4$. Fixing $\Omega_{2}$ to this critically damped point of $\kappa_{r}/4$ maximizes the imaginary component of the eigenvalue corresponding to the largest eigenvector state overlap with $\ket{n-1,g,0}$. Since this eigenvector will have the largest effective coupling to our initial error state, maximizing the imaginary component of this eigenvector will maximize the amount of dissipation that can be inherited by our initial error state, $\ket{n-1,g,0}$. Increasing $\Omega_{1}$ until the eigenvalue associated with our initial error state bifurcates, maximizes the dissipation of our error state to our code state with an optimal value of $\Omega_{1}\approx\kappa/13$. The rates used in the experiment are close to the analyzed optimal drive rates of $\Omega_{1}=\kappa_r /13,\Omega_2=\kappa_r/4$. A numerical plot of the imaginary component of this eigenvalue as a function of drive strengths is shown in Fig. 2B. For the drive rates higher than what was used in the experiment (denoted by $\times$), the eigenvalue gains a real component, making it oscillatory and hence suboptimal. 

\subsection{Driven transmon Hamiltonian}
In this section, we derive the transmon Hamiltonian under combs of off-resonant drive in the perturbative regime.  We show how four-wave-mixing (FWM) processes occurs under frequency-matching conditions, and connect the Rabi rate of the FWM transitions $\Omega_1$ and $\Omega_2$ with circuit parameters, especially its simple relationship with Stark shift, Eq.~\eqref{eq:drive_rates}.  The derivation follows the standard 4$^{th}$-order expansion of the Josephson potential in the displaced frame as in Ref~\cite{gertler2021protecting, gertler2023experimental}) which was first introduced in Ref.~\cite{leghtas2015confining}. 

We have a system containing two linear resonators and one non-linear resonator (transmon) that are capacitively coupled with each other. The linear modes are detuned from and only weakly couple to the transmon with strengths $g_{aq}, g_{rq}\ll\omega_a-\omega_q, \omega_r-\omega_q $, where $g_{aq}, g_{rq}$ are storage-transmon and reservoir-transmon coupling strengths. This weak participation of the transmon mode in the high-Q and low-Q cavity modes induces a non-linearity that allows for dispersive readout and control of the storage cavity. The Hamiltonian for this three mode system can be separated into a linear and non-linear parts such as, $\hat{H}_{tot}=\hat{H}_{l}+\hat{H}_{nl}$, where,
\begin{equation}
	\frac{\hat{H}_{l}}{\hbar} = \omega_{a}\hat{a}^\dagger\hat{a}+\omega_{q}\hat{q}^\dagger\hat{q}+\omega_{r}\hat{r}^\dagger\hat{r}
	\label{eq:ham_linear} % Use a logical label
\end{equation}
\begin{equation}
	\frac{\hat{H}_{nl}}{\hbar} = -E_{J}\left(\cos\hat{\phi} + \frac{\hat{\phi}^{2}}{2}\right)
	\label{eq:ham_nl} % Use a logical label
\end{equation}
In this hybridized Hamiltonian, $\hat{\phi}$ is written in terms of the zero point fluctuation of each mode, $\hat{\phi} = \phi_{a}(\hat{a}^\dagger+\hat{a}) + \phi_{q}(\hat{q}^\dagger+\hat{q})+\phi_r(\hat{r}^\dagger+\hat{r})$. Expanding the cosine to fourth-order and keeping only dominant photon-number preserving nonlinear interaction terms we arrive at the equation defined in the main text Eq.(\ref{eq:ham_1}). The other nonlinear terms such as storage anharmonicity given by self-Kerr $=-\frac{K}{2}\hat{a}^{\dagger2}\hat{a}^{2}$, and higher order (6$^{th}$-order) dispersive shifts $=\frac{\chi_{q}'}{2}\hat{q}^\dagger\hat{q}\hat{a}^{\dagger2}\hat{a}^2$ do not affect the correction scheme significantly and are thus omitted from the equation. For corrections on codewords with more than three Fock components, the $\chi_{q}'$ leads to slightly off-resonant pumping drives as in \cite{gertler2021protecting}.

The FWM drive combs are detuned from the transmon frequency by $\Delta_{d1-\omega_{ge}}\approx8$ GHz and $\Delta_{d2-\omega_{ge}}\approx2.7$ GHz. Due to the larger input coupling to the reservoir mode compared to the transmon, the Stark shift on the transmon from these drives is largely dominated by the displacement of the reservoir rather than direct driving. %When deriving the FWM strengths from the driven Hamiltonian, treating drives as a displacement on either of the modes gives same results. However, 
Nevertheless, we provide a general derivation assuming FWM drives couple to both the modes, 
\begin{align}
    \label{eq:ham_drive_lin}
    \frac{\hat{H}_{drive}}{\hbar}=&\sum_{n=0}^2(\epsilon_{1q,n}\hat{q}^\dagger + \epsilon_{1r,n}\hat{r}^\dagger) e^{i(\omega_{d1} - (2n+1)\chi_{gf})t} \\\nonumber +& (\epsilon_{2q}\hat{q}^\dagger + \epsilon_{2r}\hat{r}^\dagger)e^{i(\omega_{d2}+(2n+1)\chi_{gf})t} + h.c.
\end{align}
where $\epsilon_{1q/1r,n}$ is the drive strength for the first step that is constant in time and scales as $\epsilon_{1q/1r, n} = \frac{\epsilon_{1q/1r}}{\sqrt{2n+1}}$ for $n=0,1,2$ to compensate for the bosonic enhancement factor. However, $\epsilon_{2q/r}$ does not depend on $n$ since there is no change in the photon number. We are working in the regime where $\omega_{q},\omega_a, \omega_r, \omega_{d1},\omega_{d2}, \omega_{d1/d2}-\omega_{q/r} \gg \epsilon_{1q/r,n}, \epsilon_{2q/r}$, which allows us to consider the drives to be only a perturbation on the system. A unitary transformation to a displaced slowly rotating drive frame simplifies the Hamiltonian. For example, with $\hat{U}_{q, k} = e^{i\omega_{k}t\hat{q}^\dagger\hat{q}}\hat{D}_{q}(\xi_{k})e^{-i\omega_{k}t\hat{q}^\dagger\hat{q}}$ for a drive $k$ with frequency $\omega_k$ and drive strength $\epsilon_k$ the Hamiltonian can be transformed to eliminate transmon drive terms from $\hat{H}_{drive}$, when $\xi_{k}=\frac{-\epsilon_{q,k}}{ (\omega_k-\omega_q)}$. The transmon operator transforms in this frame to gain time dependence $\hat{q}\rightarrow \hat{U}_{q, k}^\dagger\hat{q}\hat{U}_{q,k}\rightarrow \hat{q} + \xi_{k}e^{-i\omega_k t}$.  
Similarly, the unitary $\hat{U}_{r, k} = e^{i\omega_{k}t\hat{r}^\dagger\hat{r}}\hat{D}_{r}(\zeta_{k})e^{-i\omega_{k}t\hat{r}^\dagger\hat{r}}$, where $\zeta_{k}=\frac{-\epsilon_{r,k}}{(\omega_k-\omega_r)}$ eliminates the drive terms coupled to the reservoir from $\hat{H}_{drive}$ and the operator gains time dependence  $\hat{r}\rightarrow\hat{r}+\zeta_ke^{-i\omega_kt}$. This displacement can be done for each of the six drive tones without loss of generality, leading to $\hat{q}\rightarrow\hat{q}+\sum_{n=0}^2 (\xi_{1,n}e^{-i\omega_{d1,n}t} + \xi_{2,n}e^{-i\omega_{d2,n}t})$ and the reservoir operator $\hat{r}\rightarrow\hat{r}+\sum_{n=0}^2 (\zeta_{1,n}e^{-i\omega_{d1,n}t} + \zeta_{2,n}e^{-i\omega_{d2,n}t})$ in the final displaced reference frame.
The operators are then substituted in the $\hat{\phi}$ of the original Hamiltonian,

\begin{align}
    \label{eq:ham_phi_zpf} 
    \hat{\phi}=&\phi_{a}\hat{a} + \phi_{q}\hat{q}+\phi_r\hat{r} + \sum_{n=0}^2 [ (\phi_q\xi_{1,n}+\phi_r\zeta_{1,n}) e^{-i\omega_{d1,n}t} \\\nonumber &+(\phi_q\xi_{2,n}+\phi_r\zeta_{2,n})e^{-i\omega_{d2,n}t}] + h.c 
\end{align} 

This can be simplified by replacing the weighted sum of displacements with a constants $\phi_q\beta_{1,n} = \phi_q\xi_{1,n}+\phi_r\zeta_{1,n}$ and $\phi_q\beta_{2,n} = \phi_q\xi_{2,n}+\phi_r\zeta_{2,n}$.
 
Finally, we perform another unitary transformation on the Hamiltonian,
\begin{equation}
	\hat{U} = e^{i\omega_qt\hat{q}^\dagger\hat{q}}e^{i\omega_at\hat{a}^\dagger\hat{a}}e^{i\omega_rt\hat{r}^\dagger\hat{r}}
	\label{eq:Unitary_s5} 
\end{equation}
The operators in this frame are $\hat{a}\rightarrow e^{-i\omega_at}\hat{a}, \hat{q}\rightarrow e^{-i\omega_qt}\hat{q}, \hat{r}\rightarrow e^{-i\omega_rt}\hat{r}$ and the Hamiltonian changes as $\hat{H}_{rot}\rightarrow \hat{U}\hat{H}\hat{U}^\dagger + i\hbar \dot{\hat{U}}\hat{U}^\dagger$ which cancels $\hat{H}_{l}$. We can then do a fourth-order expansion of the cosine to   
\begin{align} 
     \label{eq:ham_rot}
	\frac{\hat{H}_{rot}}{\hbar} = & - \frac{\alpha_{q}}{2}\hat{q}^{\dagger}\hat{q}^{\dagger}\hat{q}\hat{q}  -\chi_{ge}|e \rangle\langle e| \hat{a}^{\dagger} \hat{a} \\\nonumber
    -&\chi_{gf}|f \rangle\langle f|\hat{a}^{\dagger}\hat{a} - \frac{K}{2}\hat{a}^{\dagger 2}\hat{a}^{2} - \chi_{qr} |e \rangle\langle e| \hat{r}^{\dagger}\hat{r}
    \\\nonumber
    -&\sum_{n=0}^2[\frac{E_{J}}{2}\phi_q^3\phi_a\beta_{1,n} e^{i(\alpha_q +(2n+1)\chi_{gf})t}\hat{q}^{\dagger2}\hat{a}^{\dagger} 
    \\\nonumber +&\frac{E_{J}}{2}\phi_{q}^{3}\phi_{r}\beta_{2,n} e^{-i(\alpha_q+(2n+1)\chi_{gf})t}\hat{q}^{2}\hat{r}^{\dagger}] + h.c.   
\end{align}

The drives also produce a stark shift on the transmon, which was absorbed in its frequency before changing the frame in the above equation. These stark shift terms can be calculated from the fourth-order terms of the cosine expansion. The following are the non-rotating and slowly rotating terms from the expansion that produce the stark shift on the transmon.  

\begin{align}
    \label{eq:ham_stark}
    \frac{\hat{H}_{stark}}{\hbar} = &-E_J\phi_{q}^4\sum_{n=0}^{2}(|\beta_{1,n}|^{2}+|\beta_{2,n}|^{2})\hat{q}^{\dagger}\hat{q}
    \\\nonumber
    &- E_J\phi_q^4\sum_{k=0}^{2}\sum_{\substack{l=0 \\ l\neq k, l>k}}^{2}[\beta_{1,k}\beta_{1,l}^*e^{i2(k-l)\chi_{gf}t} \\\nonumber &+ \beta_{2,k}\beta_{2,l}^*e^{i2(l-k)\chi_{gf}t}]\hat{q}^\dagger\hat{q}  
\end{align}

Ignoring the slow-rotating cross terms and writing only the static shift in transmon frequency in terms of the known parameters $\chi_{qr}, \chi_{ge}, \alpha_q$ we get the following, 
\begin{equation}
    \Delta_{ss} = 2\alpha_{q}\sum_{n=0}^{2}(|\beta_{1,n}|^{2}+|\beta_{2,n}|^{2})
	\label{eq:ham_s7_stark2} 
\end{equation}

The drive rates to first order are just coefficients of $\hat{q}^{\dagger2}\hat{a}^\dagger$  and $\hat{q}^{2}\hat{r}^\dagger$ from the Eq.~\ref{eq:ham_rot} with the bosonic factor $\sqrt{2}$ for adding two excitations to the transmon $|g\rangle\rightarrow|f\rangle$ and $\sqrt{2n+1}$ where $ n=0,1,2$ for cavity excitation.     
\begin{align}
	\Omega_{1,n} =& \frac{E_J}{2}\phi_q^3\phi_a\beta_{1,n}\sqrt{2(2n+1)}%=\sqrt{2(2n+1)\chi_{ge}\alpha_q}\beta_{1,n}
    =\sqrt{(2n+1)\chi_{ge}\Delta_{ss1}}    \nonumber\\
    \Omega_{2,n} =&  \frac{E_J}{2}\phi_q^3\phi_r\beta_{2,n}\sqrt{2(2n+1)}%=\sqrt{2(2n+1)\chi_{qr}\alpha_{q}}\beta_{2,n} 
    = \sqrt{(2n+1)\chi_{qr}\Delta_{ss2}}
	\label{eq:drive_rates} % Use a logical label
\end{align}

The $\beta_{1,n}=\sqrt{\frac{\Delta_{ss1}}{2\alpha_q}}{}$ and $\beta_{2,n}=\sqrt{\frac{\Delta_{ss2}}{2\alpha_q}}$ are substituted above to get the drive rates in terms of Stark shift on transmon from each tone of the two drive pumps. These calculated drive rates $\Omega_{1,n}, \Omega_{2,n}$ for one transition $n=0$ is listed in Table.~\ref{tab:drive_params}.   

\subsection{Spurious qubit excitations under weak off-resonant drives}

\begin{figure*}[tbp] % Do not use \begin{figure*}
	\centering	
    \includegraphics[width=17cm]{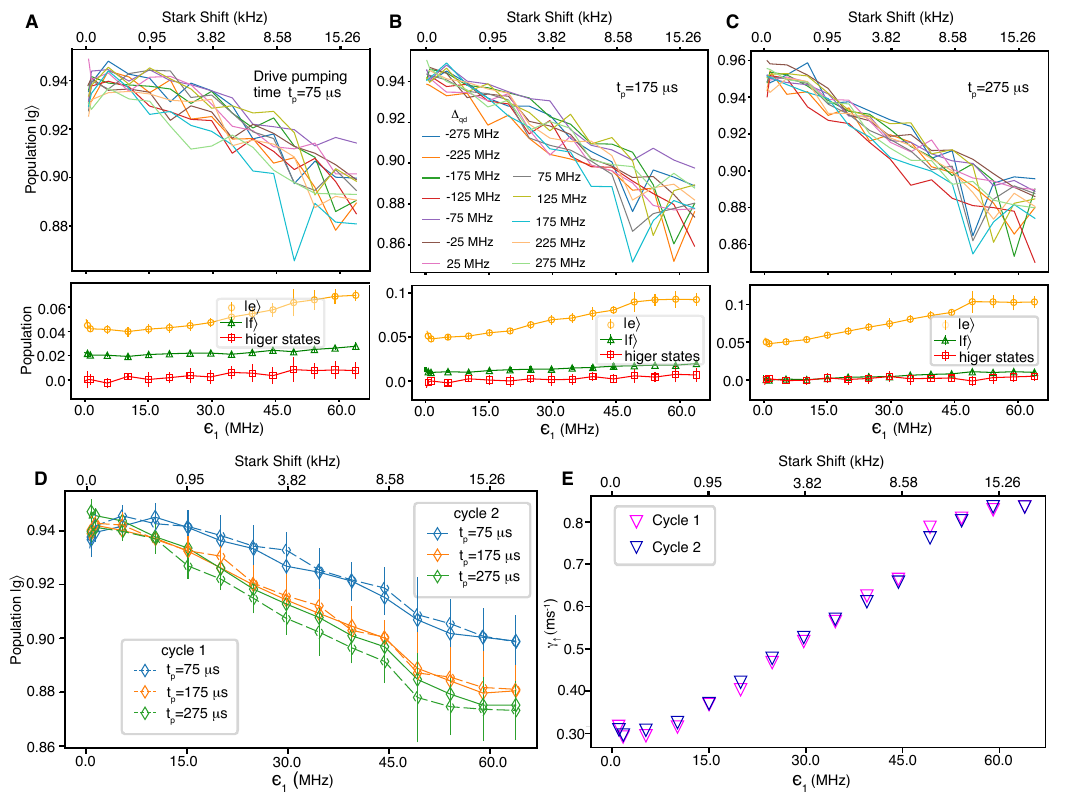} 
	\caption{\textbf{Transmon excitations by a single-tone off-resonant drive.}
	(\textbf{A}) The $|g\rangle-$state populations of the transmon driven with a single parameter ($\epsilon_1$) sweep of an off-resonant tone that is applied for $t_p=75\,\mu$s. Each curve represents a measurement repeated for a drive frequency detuned from the FWM transition, $\omega_{d1}$. The subplot below shows the $\ket{e}, \ket{f}$ and higher excited state populations, averaged across measurements with different drive frequencies. In (\textbf{B}), (\textbf{C}), the measurement is repeated for different drive pumping times $t_p=175\,\mu$s, $t_p=275\,\mu$s, respectively. The shared legend for all the figures is shown in plot (\textbf{B}). (\textbf{D}) The $\ket{g}-$state population is averaged across drive frequencies for the three pump times. The error bars represent the spread of the data across drive frequencies. The cycle 1 and cycle 2 are measurements from two iterations of the same experiment. (\textbf{E}) The excitation rate $\gamma_{\uparrow}$ is analyzed from population data for different drive amplitudes $\epsilon_1$. The Rabi rate $\Omega_1=55$ kHz requires relatively weak drive, $\epsilon_1\approx 26$ MHz that yields a Stark shift $\Delta_{ss}=3$ kHz and corresponds to $\gamma_{\uparrow}\approx 0.5$ ms$^{-1}$. %, which is a 66\% increase from the baseline. 
    }
	\label{fig:qub_temp_1} 
\end{figure*}

A main concern of continuous AQEC with driven dissipation is that the always-on off-resonant drive tones may cause spurious excitation of the transmon, which in turn will dephase the logical qubit. To investigate whether this process may occur for the drive tones used in the AQEC experiment reported in the main text, we apply intentionally-detuned PReSPA drives to the initial equilibrium state of the system (very close to $\ket{0g0}$) for a long period of time, and measure the excited state population $P_e$ of the transmon.  Due to the lack of high-fidelity shot-single measurements, we use the $\ket{e}-\ket{f}$ Rabi oscillation to measure $P_e$~\cite{jin2015thermalresidue, geerlings2013drivenreset}. Detuning of 10's to 100's of MHz was introduced to avoid any resonant excitations instigating the four-wave mixing (FWM) processes. We find steady-state $P_e=3.8\%$ under the continuous pumps at experimental settings during the same periods when the data presented in the main text was acquired, which is higher than the $P_e=1.7\%$ when system is in equilibrium. The transmon $T_1$ was not degraded under these continuous detuned drives. This additional transmon population can be converted to an effective additional heating rate rate $\Delta\gamma_\uparrow$ of 0.42 ms$^{-1}$. 

We performed a more systematic measurement of this drive-induced qubit excitation as a function of the drive strength $\epsilon_1$ and $\epsilon_2$ (which can be converted to FWM Rabi rates $\Omega_1$ and $\Omega_2$) in a subsequent cooldown of the same device with slightly different coupling parameters. The transmon $T_{1ge}\approx$ 170 $\mu$s, $T_{1ef}\approx$ 75 $\mu$s in this cooldown is longer than previously measured, allowing for greater resolution in detecting a small heating effect.  We further used a driven-dissipative reset sequence to initialize the transmon in $\ket{g}$ by a driving the system from $\ket{0e0}$ or $\ket{0f0}$ to $\ket{0g1}$ from where it spontaneously decays, which increased the throughput of the experiment.  We further applied shot-to-shot sweeping of the drive amplitude to distinguish between instantaneous driven response of the system and any long-term cumulative impact on the environment (such as heating up off-the-chip microwave components of the system, which are expected to react in timescales of seconds or longer).  

In Fig.~\ref{fig:qub_temp_1}, we present the measured transmon population in $\ket{g},\ket{e},\ket{f}$ or higher states following three different pumping times $t_p=75$ $\mu$s, 175 $\mu$s, 275 $\mu$s.  One single-frequency pumping tone is applied at various detunings from $\omega_{d1}=11.5$ GHz.  In the top panel of A, B, C, each curve corresponds to one independent measurement where the drive amplitude $\epsilon_1$ is swept along the $x$-axis from left to right from shot to shot, averaged over many rounds of the sweep.

In the experiment, after pumping for time $t_p$ for each $\epsilon_1$, we obtain readout values ($D_1, D_2, D_3, D_4$) from four different iterations of the measurements using the following transmon pulse sequences, 
\begin{enumerate}
    \item no transmon pulse,
    \item $\ket{e}\leftrightarrow \ket{f}$-pulse followed by $\ket{g}\leftrightarrow \ket{e}$-pulse,
    \item  $\ket{g}\leftrightarrow \ket{e}$-pulse followed by another $\ket{e}\leftrightarrow \ket{f}$ followed by $\ket{g}\leftrightarrow \ket{e}$-pulse,
    \item  $\ket{g}\leftrightarrow  \ket{e}$
\end{enumerate} 
which give four equation $gA+eB+fC+R=D_1$, $fA+gB+eC+R=D_2$, $fA+eB+gC+R=D_3$ and $eA+gB+fC+R=D_4$, where $g,e,f$ are the populations in respective states.  $A,B,C$ are the readout values (in arbitrary units) for transmon in $\ket{g}, \ket{e}, \ket{f}$ state, and $R$ is a small residual signal from potential higher-excited state occupancy.  We do not generally impose constraints of $g+e+f=1$ since it is possible that the drive may excited the transmon to high excited states.  At small $\epsilon_1 \approx 0$, after sufficient $t_p$ after reset, the system is assumed to be in equilibrium with population only in $\ket{g},\ket{e}, \ket{f}$ states, which we use for a self-consistent calibration of $A, B, C$ throughout the duration of the experiment. %For the three different pump times $t_p$ we assume different $f$-population at equilibrium and solve for the four variables $A, B,C, e$ with the constraint $g=1-e-f$. 
Using the calibrated $A,B,C$ values, %from this solution of equations, 
the data ($D_1,D_2, D_3, D_4$) is used to solve for $g,e,f$ and $R$ %A least-square optimization for the above equations with an additional parameter $+hH$ to the left-hand side 
giving the population values as plotted in Fig.~\ref{fig:qub_temp_1} (where the ``higher-state'' population is computed as $1-g-e-f$).

\begin{figure*}[tbp] % Do not use \begin{figure*}
	\centering
	\includegraphics[width=17cm]{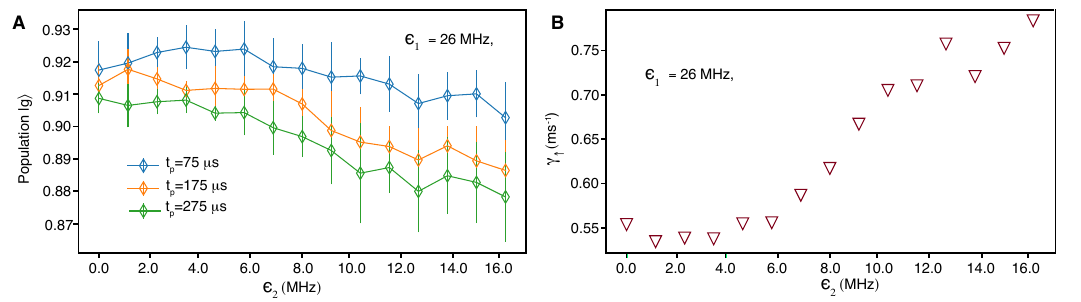} % for an image file named example_figure.*
	% Pick an appropriate width for the size of the image

	% Captions go below figures
	\caption{\textbf{Transmon excitations by two combs of off-resonant drives.} The two combs, each with three tones spaced by $2\chi_{gf}$, are placed at 100's of MHz detuned from $\omega_{d1}$ and $\omega_{d2}$ to mimic the FWM pumping strength under the AQEC condition. \textbf{(A)} $\ket{g}-$state population as a function of $\epsilon_2$, which is averaged over 12 different drive frequencies detuned from $\omega_{d2}$ and is presented for three different pump times. The measurement has one drive comb at fixed strength $\epsilon_1=26$ MHz and fixed detuning from $\omega_{d1}$. \textbf{(B)} The $\gamma_{\uparrow}$ rate obtained from analyzing the population data of \textbf{(A)}. The drive strengths that produce the Rabi rates $\Omega_{1}, \Omega_{2}$ used in the experiment are $\epsilon_1\approx26$ MHz,  $\epsilon_2\approx11.5$ MHz, respectively, which would produce a $\gamma_{\uparrow}\approx0.7$ ms$^{-1}$.    
	}
	\label{fig:qub_temp_2} % give each figure a logical label name
\end{figure*}

As seen in Fig.~\ref{fig:qub_temp_1}(\textbf{A, B, C}), with increasing $\epsilon_1$ the transmon is driven out of the $\ket{g}$ state. In the sub-panels, we plot the $\ket{e},\ket{f}$ and higher $\ket{h}$ state populations, as calculated from the above analysis. The populations mostly do not depend on the detuning of the drives from the FWM frequency allowing us to plot average of all of them with error bars representing the standard deviation of the spread for different frequencies in Fig.~\ref{fig:qub_temp_1}D. The pump time of $t_p=275\,\mu$s is the longest we have measured, at which the transmon approaches a steady state. 

It should be noted that our readout produces some high-excited state beyond $\ket{f}$ due to measurement induced state transition~\cite{dumas2024mist}, which cannot be efficiently reset to $\ket{g}$ and is responsible for our $\ket{f}$-state population of approximately 2\% at a wait time $t_p=75$ $\mu$s after the transmon reset even at neglible drive power. From the data, we observe that the $\ket{f}$ state population consistently goes down as $t_p$ increases from 75 $\mu$s to 275 $\mu$s even at the highest drive power.  The dominant effect of increasing drive power appears to be populating $\ket{e}$-state rather than producing high-excited states which then gradually cascade down the transmon levels over time. 

We believe the higher-excited state population at higher drive amplitude is caused by relatively instantaneous effects within each reset-measurement cycle (of 100 - 300 $\mu$s) rather than any cumulated heating over much longer time scales.  This is ensured by our measurement mode of sweeping up $\epsilon_1$ from shot to shot.  We have also carried out measurements where we sweep $\epsilon_1$ down, yielding very similar results.  (Some minor differences are observed for the first points after a ``loop-over" which we attribute to a slightly different reset fidelity.  Even if it indicates a cumulative heating effort, it poses an upper bound to the memory timescale to $\lesssim$1 ms.)

Motivated by the above observation, we analyze the spurious excitation based on a simple instantaneous heating model. 
We assume that the ratio between the excitation rates and relaxation rates, $r=\gamma_\uparrow/\gamma_{\downarrow}$, between the neighboring transmon states, is the same for $\ket{g}-\ket{e}$ and $\ket{e}-\ket{f}$ transitions (neglecting the minor transition frequency differences in the detailed balance relation) but is a function of the drive amplitude, i.e.~$r=r(\epsilon_1)$. %remains the same which becomes a parameter that we sweep to fit the time domain population for different FWM amplitudes. 
We did not observe any appreciable degradation to the transmon $T_1$ time in the presence of the drives, hence we assume the down-transition rates do not change under drives. Then we can model the evolution of the transmon state based on a rate matrix with a single parameter $r$, which we use to fit the measured qubit population data at four different pumping times ($t_p=75$ $\mu$s, 175 $\mu$s, 275 $\mu$s plus the initial state $t_p=0$), independently for each $\epsilon_1$. The fit result for $r$ is converted to the $\ket{g}\rightarrow\ket{e}$ transition rate $\gamma_\uparrow=rT_{1ge}$ and presented in Fig.~\ref{fig:qub_temp_1}E.  We see that this rate increases approximately linearly with the pump amplitude, except for perhaps a very small curved region near $\epsilon_1=0$ where the Stark shift is well below 1 kHz.  We acquire data in two cycles which show similar results over varying amplitudes. The baseline here at equilibrium is 0.3 ms$^{-1}$.  % which is similar to the rate extracted by barely measuring the $P_e$ in the experiment.

In Fig.~\ref{fig:qub_temp_2}, we plot the analayzed population data for a comb of 3 drive tones near $\omega_{d1}$ with fixed amplitude and a comb of 3 drive tones near $\omega_{d2}$ with varying amplitude. This measurement is also carried out at different detunings and averaged over different detuning frequency values. The amplitude of the first drive comb is calibrated to match the AQEC condition presented in the main text.  %so we need to calibrate the $A, B, C$ values slightly differently by assuming correct $\ket{e}$-state population at $\epsilon_2\approx0$. 
The AQEC condition used in the main text corresponds to $\epsilon_2$ = 11.5 MHz which from the figure gives $\gamma_{\uparrow}=0.7$ ms$^{-1}$ that is about $\Delta\gamma_{\uparrow} = 0.4$ ms$^{-1}$ above the baseline.  

Despite the very different qubit $T_{1}$ time and a different attenuator and filter configuration in this cooldown, we find a very similar drive-induced excitation rate to what was coarsely measured during the AQEC experiment.  This coincidence, together with the lack of memory effect, suggests that this apparent heating rate may be intrinsic to the driven transmon rather than extrinsic.  However, the near-linear relationship is particular surprising as any multi-photon off-resonance excitation mechanism should have a quadratic or higher power-law dependence on drive amplitude.  In fact, current understanding in DUST~\cite{dumas2024mist} suggests the existence of turn-on thresholds that is much higher than the regime explored here.  The mechanism of the unexpected DUST observation in this low-power pumping regime with multi-millisecond transition rates will therefore require theoretical understanding of weaker and more subtle effects in the driven circuit QED systems.

\subsection{Comparing continuous passive QEC with active QEC in circuit QED}

\begin{table*}[t!]
    \centering
    \caption{\textbf{Error budget comparison of active versus passive QEC. }
        Here we present how each error contributes to the loss of fidelity in the logical code space in the measurement-based vs passive QEC in the binomial code. We show an approximate analytic scaling on the left side of the columns and an associated calculated error rate \textit{using the device parameters from this AQEC experiment} on the right side of the columns.  For computing the error budget in active QEC, we consider two cases: 1) A base case with the same $\bar{n}=3$ binomial code and correction rate $\kappa_{cor}=\frac{2}{\tau_{cyc}}=(4$ $\mu\text{s})^{-1}$ as in our AQEC experiment for a more direct comparison, and 2) An example case with the $\bar{n}=2$ kitten code and $\kappa_{cor}=(8$ $\mu\text{s})^{-1}$ optimized for overall logical error rate.  In either case, we have applied the performance metrics of the ancilla readout, reset, and recovery unitary from Ref.~\cite{ni2023beating} for calculating error rates: $\tau_{ex}$ = 1.3 $\mu$s is the total ancilla excitation time after a positive syndrome measurement; $\epsilon_{meas}=1.4\%$, $\epsilon_{j}=3.9\%$ and $\epsilon_{nj}=1.4\%$ are the readout, jump correction and no-jump correction infidelities, respectively. %The $\cdots$ signify error rates that are either not applicable to the QEC scheme or have negligible error contributions. 
        The notation $f(x)$ denotes a functional dependence whose detail is beyond the scope of the discussion.}
    \label{tab:error_budget_comparison}
    \renewcommand{\arraystretch}{1.2} % makes it a bit more readable
    \vspace{0.5em}
    \begin{tabular}{|c|c|cc|c|c|}
    %    \vspace{}
        \hline
        \multirow{3}{*}{Error sources} & 
        \multicolumn{3}{c|}{Active measurement-based QEC} & 
        \multicolumn{2}{c|}{Continuous AQEC} \\
        \cline{2-6}
         & \multirow{2}{*}{Analytic scaling} & \multicolumn{2}{c|}{Error rate (ms$^{-1}$)}  & \multirow{2}{*}{Analytic scaling}  & Error rate (ms$^{-1}$)\\
         &      &  $(\bar{n}=3)$  &  $(\bar{n}=2$, opt.)   &    & ($\bar{n}=3$)  \\
     %   \hline
        \hline
        Double photon loss 
        & $\bar{n}^{2}\kappa_a^2/\kappa_{cor}$
        & 1.8
        & 1.8
        & $\bar{n}^{2}\kappa_a^2/\kappa_{cor}$ 
        & 1.8 \\
        \hline
        \begin{tabular}[c]{@{}l@{}}\;\, Ancilla relaxation \\ $|e\rangle \rightarrow |g\rangle$ or $|f\rangle \rightarrow |e\rangle$\end{tabular}
        & $\bar{n}\kappa_a(\tau_{ex}/T_{1ge})$
        & 0.5
        & 0.33
        & $(\bar{n}\kappa_a)/(5\kappa_{cor}T_{1ef})$
        & 0.6 \\
        \hline
        Unwanted corrections  
        & $\frac{1}{2}\epsilon_{meas}\kappa_{cor}$
        & 1.3
        & 0.65
        & $4\kappa_{cor}^3/\chi_{gf}^2$
        & 0.4 \\
        \hline 
        Imperfect recovery unitary  
        & $\frac{1}{2}\epsilon_{nj}\kappa_{cor}+\bar{n}\kappa_{a}\epsilon_{j}$ 
        & 2.6
        & 1.4
        & $\kappa_a f(\bar{n})$ 
        & 1.3 \\
        \hline
        Storage mode nonlinearity  
        & $\frac{1}{6}(K/\kappa_{cor})^2\bar{n}\kappa_{a}$ 
        & 0.1 
        & 0.27
        & $\frac{1}{6}(K/\kappa_{cor})^2\bar{n}\kappa_{a}$ 
        & 0.1 \\
        \hline
        \begin{tabular}[c]{@{}l@{}}Ancilla spurious excitation \\ $\qquad\qquad |g\rangle \rightarrow |e\rangle$ \end{tabular}
        & $\gamma_{\uparrow 0}$
        & 0.3
        & 0.3
        & $\gamma_{\uparrow 0}+f(\Omega_1, \Omega_2)$%\Delta\gamma_\uparrow$
        & 0.7 \\
        \hline 
        Ancilla Dephasing  
        & $(\pi\bar{n}\kappa_a )/(\chi_{ge}T_{\phi})$ 
        & 0.2 
        & 0.2
        & $(\bar{n}^{2}\kappa_a^2)/(\kappa_{cor}\Omega_{2}^2T_{\phi}^{2})$
        & $<0.01$ \\
     %   \hline
     %   \hline
     %   Total  
     %   & 
     %   & 6.3
     %   & 4.85
     %   & 
     %   & 4.9 \\
        \hline
    \end{tabular}
\end{table*}

In this section of the supplementary we do a comparative analysis of the errors in the two approaches i.e., active measurement based and passive QEC for binomial qubits with a transmon ancilla.  %In order to make the encoding of quantum information hardware efficient, we use bosonic modes in microwave cavities whose
The dominant error channel, single photon loss at rate $\kappa_a$, induces both jump and no-jump evolution % in the bosonic logical qubit 
that need to be corrected. In active measurement-based corrections, such as Ref.~\cite{ni2023beating} which we will extensively use as our comparison, a correction unitary was applied based on a measurement outcome that corrected for jump or no-jump errors accumulated over a finite time $t_{wait}$. In our protocol however, we engineer a dissipative scheme to correct for the parity-flip error from the jumps. There remains a distortion effect in the codeword coefficients from the jump or no-jump evolution at finite $\bar{n}$ that would ultimately need a four-photon dissipative process~\cite{vanselow2025dissipatingquartets} to fully stabilize.  To mitigate this uncorrected effect, we used a larger photon number, $\bar{n}=3$, compared to $\bar{n}=2$ in Ref.~\cite{ni2023beating}.  Apart from the jump and no-jump errors, the bosonic logical states are also affected by ancilla errors and the induced nonlinearities in the harmonic oscillator modes which are not corrected for in either of the correction schemes. As a pedagogical comparison, we will address each error avenue and compare its mechanism and magnitude of error for the two different schemes, with a full summary listed in Table.~\ref{tab:error_budget_comparison}. 

\subsubsection{Double photon loss errors}
Incurring a second photon-loss error before the first is corrected is the canonical 2nd-order error that is uncorrected in a 1st-order QEC implementation.  Its rate is the result of direct competition between the raw error rate $\bar{n}\kappa$ and the rate of error correction $\kappa_{cor}$. For a binomial code, a double-photon-loss error causes an effective bit flip error while leaving the parity unchanged and is thus uncorrectable in either passive or active QEC. The double photon loss rate is $\bar{n}^2\frac{\kappa_{a}^{2}}{\kappa_{cor}}$ for both correction schemes. This expression can be derived from Poisson statistics in the limit of $\kappa_{cor}\gg\kappa_{a}$. We have defined the correction rate ($\kappa_{cor}$) as the inverse of one half an active QEC cycle time $\tau_{cyc}$ (which is often chosen to be slower than the speed limit of parity tracking), $\kappa_{cor}=2/\tau_{cyc}$. In the quantitative example comparison of Table~\ref{tab:error_budget_comparison}, we choose $\tau_{cyc}=2/\kappa_{cor}=2\tau_{cor}=8$ $\mu$s for the active QEC, which is twice the characteristic time of our experimental AQEC ($\tau_{cor}=1/\kappa_{cor}= 4$ $\mu$s, which we will also refer to as the effective passive ``cycle time'') so that the baseline 2nd-order error rate is held equal. The heuristic picture for this factor of two difference is that the continuous correction only starts after a photon loss event, thus the system is susceptible to the second photon loss error for the duration of the characteristic correction time $1/\kappa_{cor}$.  The active QEC cycles run on a schedule uncorrelated with the loss events. Thus, on average, the time interval between the first loss event and the next correction operation is $\tau_{cyc}/2$, motivating our definition of $\tau_{cyc}/2=\tau_{cor}$.  
%the system is susceptible for half of the the first loss happens in the first half of the wait time, and the second loss happens in the remaining half of the time. This gives a narrower time frame for the second loss to occur, on average, in the active scheme. For the same correction rate, the double photon loss has a lesser effect in the active correction protocol than the continuous AQEC.  
%In the quantitative example comparison of Table~\ref{tab:error_budget_comparison}, we choose $\tau_{cor}=1/\kappa_{cor}=8$ $\mu$s for the active QEC, which is twice the characteristic time of our experimental AQEC ($\tau_{cor}=1/\kappa_{cor}= 4$ $\mu$s, which we will also refer to as the effective ``cycle time'') so that the baseline 2nd-order error rate is held equal. 

\subsubsection{Unwanted corrections}
The interactions in both schemes leading to correction come at a cost, since they also introduce associated errors. In the passive scheme, the combs of four-wave mixing tones can cause detuned transitions out of the logical codespace. In the measurement-based scheme, incorrect readout assignment can lead to an incorrect unitary correction being applied, thus scrambling the information and driving the system out of the logical codespace. 
%These errors are similar in behavior, in that they both arise from an incorrect application of a correction operation, and thus can be compared on equal footing.

For the error rate of a detuned transition in the continuous scheme, we can refer to the $3\times3$ matrix describing the cascaded dissipative dynamics, Eq.~\eqref{eq:diss_matrix}, with a detuning term of $\chi$, 
\begin{equation}
\hat{H}_\text{detune}=\begin{pmatrix}
0 & \Omega_1 & 0\\
\Omega_1 & \chi & \Omega_2\\
0 & \Omega_2 & -i\kappa/2\\
\end{pmatrix}
\end{equation}
%where $\Omega_{1}, \Omega_{2}, \kappa_{r}, \chi_{gf}$ are drive rate for the first tone, second tone, reservoir loss rate and $gf-$dispersive coupling strength respectively.
In the limit of $\kappa\gg\Omega_2\gg\Omega_1$, the imaginary component of the eigenvalue of interest yields a detuned transition rate of $\frac{\Omega_1^2\Omega_2^2}{\chi^2\kappa}$ which approaches $\approx \frac{\Omega_1^3}{\chi^2}$
as the drives approach the critical damping regime. This shows the clear benefit that an increase in $\chi$ provides to this error. 
%The $\kappa_{cor}^3$ scaling is important here as when contrasted with the $1/\kappa_{cor}$ scaling of the double photon loss, it depicts a clear relation to use to optimize the drives to minimize the total losses of these two avenues. 
It should be noted that this detuned transition can be further mitigated to a large degree in a hybrid active/passive QEC scheme~\cite{de2022error,lachance2024autonomous} by using cascaded pulses with spectrally narrow envelopes instead of continuous ones. If the first pump tone uses a gaussian envelope with a temporal width of $\sigma_{q}$, the spectral width will also follow a gaussian dependence where the amplitude at our detuned transition will be $\propto e^{-\chi^{2}\sigma_{q}^2/2}$. By choosing a gaussian pulse with appropriate width, one can minimize destabilization of the code state from these detuned transitions to a great extent. %While this would provide a large benefit for the experiment at hand, the continuous protocol was chosen since the drive strengths needed for the cascaded pulse approach were significantly larger and there was an experimental limitation of the maximum drive strengths that could be played from the current microwave setup.

%For the measurement-based scheme, the errors arising from interactions with the system can be grouped into either measurement errors or errors from imperfect correction of the state.
In the active scheme, this error comes from a measurement infidelity ($\epsilon_{meas}$). Ref.~\cite{ni2023beating}, using the measurement-based approach, had on average $\epsilon_{meas}\sim 1.4\%$ measurement infidelity. Therefore, frequent application of active QEC cycles are often undesirable. %The fundamental limit from photon shot noise on readout fidelity is very small, making these error rates nowhere near a fundamental limit, so 
There is substantial room for improvement on readout fidelity, %a large amount of improvement that can be done here from an engineering standpoint. Improving the microwave setup and ancilla coherence lifetimes would increase this readout fidelity, for example. The details of optimizing readout are beyond the scope of this discussion, so this fidelity value will be used as is in the current comparison.
but it faces strong limitations in measurement-induced state transitions, % and decoherence of the ancilla in a sophisticated ancilla-cavity join unitary operation, the details of 
which is beyond the scope of this discussion. The passive case has a lower error rate for this pathway with our current experimental parameters (0.4 ms$^{-1}$ vs. 1.3 ms$^{-1}$, as in Table~\ref{tab:error_budget_comparison}). Since the measurement-based scheme error rate is $\propto \kappa_{cor}$, while the continuous approach is $\propto \kappa_{cor}^3$, the passive scheme benefits greater from a reduced $\kappa_{cor}$ as storage lifetimes increase. %, due to the , due to the scaling difference. Thus, the passive case only gets relatively better as longer lived storage modes are used.
%The error probability per correction cycle for the continuous protocol can be interpreted as $\kappa_{cor}^2/\chi_{gf}^{2}$, which is $0.06\%$ in the current experiment. Comparing this to the measurement interaction error of $\sim1.5\%$ for the active scheme, the continuous approach has an advantage here. %This is quoted for storage lifetimes of the order of $\sim100\mu s$ or longer. Both approaches leave room for improvement as engineering challenges, but 
%In most experimentally relevant parameter ranges with current technology, the continuous approach will have significantly lower interaction error rates.

\subsubsection{Ancilla errors}
The ancilla is an integral part of a QEC protocol. % and ancilla errors may lead to a partial or complete failure of the error correction process in both cases. 
The errors on the ancilla, such as $\ket{e}\rightarrow \ket{g}$ decay, pure dephasing, or spurious $\ket{g}\rightarrow \ket{e}$ excitation (with rate $\gamma_{\uparrow}$) may lead to a loss of logical information. Since the ancilla is utilized in both schemes a bit differently, the rate at which it affects the logical states differ. 

The errors from the heating rate $\gamma_{\uparrow}$ give a constant error rate, irrespective of the correction rate, in both protocols. However, in the passive case, due to the external drives, the effective $\gamma_{\uparrow}$ is higher than the bare un-driven rate $\gamma_{0\uparrow}$, giving a benefit to the active case. Furthermore, when the corrections are done in cycles rather than continuously, it is in principle possible to decouple the transmon during wait times~\cite{valadares2024demand} using a tunable ancilla to mitigate the dephasing effect of excitations during the idle wait times. In the continuous AQEC, the correction happens stochastically, and thus we cannot completely decouple the transmon from the oscillator mode. % as there is no clear distinction for when the ancilla is "needed" for correction or not. 
A possible approach for the passive scheme to mitigate this issue is to use strong drives that cancel $\chi_{ge}$, as done in~\cite{rosenblum2018faulttolerant}, to protect the bosonic state from dephasing from the transmon $\ket{g}-\ket{e}$ transitions (while our PReSPA uses the $\ket{f}$-state). As already discussed, we observe that with strong drives the $\gamma_{\uparrow}$ rate increases, thus any drives in addition to the four-wave mixing tones used for correction are going to increase it more and this could pose experimental challenges.

The second ancilla error channel is the $\ket{e}\rightarrow\ket{g}$ or $\ket{f}\rightarrow\ket{e}$ decay, which causes cavity logical dephasing.  It is relevant only for the time in which the ancilla is in the excited state during the correction process. In the passive case, the ancilla is in the $|f\rangle$-state for $\sim20\%$ of the correction time $t_{cor}$ under critical damping dynamics. Thus, the probability of ancilla decay for each correction is roughly $\frac{t_{corr}}{5T_{1}}$. The active protocol is susceptible to ancilla decay starting from parity mapping and ending at ancilla reset.   %, the $T_{1}-$decay may also cause an incorrect state assignment that is followed by application of the incorrect unitary. 
As the cavity lifetimes become longer we could perform fewer correction cycles over time in the active case and thus reduce %the effects of ancilla decay on logical information. Therefore, using a slower correction rate lowers 
the ancilla-$T_1$-induced errors in the active scheme. The situation with continuous AQEC is more complicated.  If we assume the use of critically-damped cascaded dissipation (which minimizes unwarranted corrections and heating), a slower correction implies the use of slower reservoir decay rate $\kappa_r$.  This leaves the ancilla in the excited state for longer and increases the probability of ancilla decay per correction, leaving ancilla-$T_1$-induced logical error rates a constant.  However, if we assume fixed $\kappa_r$ and go deeper into the overdamped regime as cavity lifetime improves (thus accepting a steeper reduction of correction rate than the reduction of FWM pumping amplitude), slower correction reduces the excitation probability of the ancilla during the correction, yielding the same benefit as in the case of active QEC. 

The third ancilla error is dephasing, with timescale $T_{\phi}$.  %, which slows down the correction in both cases. 
The parity mapping in the active scheme needs to be fast to avoid dephasing errors that can lead to incorrect mapping followed by an incorrect correction unitary. In the passive scheme, the decoherence in the $|g\rangle-|f\rangle$ subspace merely slows the first step of correction without causing logical error. 
%To understand this, note that the dephasing errors manifest as a stochastic change in the ancilla frequency over time. If we have phase flip rates at a rate of $1/T_{\phi}=\Gamma_{\phi}$ then this represents the ``spread'' in frequency of ancilla. 
To a good approximation, the ancilla frequency changes by $\approx \Gamma_{\phi}$ over time, acting as an effective detuning. In the passive case, this results in a slower correction rate by a factor of $\approx (1-\frac{\Gamma_{\phi}^{2}}{\Omega_{2}^{2}})$ which will increase the double photon loss rate by $\approx\bar{n}^2\frac{\kappa_{a}^2}{\kappa_{cor}}\left(\frac{\Gamma_{\phi}^{2}}{\Omega_{2}^{2}}\right)$ in the limit where $\Gamma_{\phi}\ll \Omega_{2}$. For most reasonably long $T_{\phi}$ values ($\gtrsim 10\mu s$), this will be a negligible effect. In the active scheme, ancilla dephasing errors will occur during the error mapping period ($t_{map}\approx\pi/\chi_{ge}$), resulting in a flipped ancilla state and thus an incorrect assignment, resulting in an error rate of $\approx\frac{\pi\bar{n}\kappa_a}{\chi_{ge}T_{\phi}}$. This is a relatively small error for reasonably long $T_{\phi}$ values. It should be noted that this error can be mitigated by performing multiple syndrome measurements and majority voting.  However, since the error rate is small, the error overhead incurred by three or more measurements is larger than the mitigated error from this dephasing effect in most cases so it is usually not practical. Overall, both schemes suffer little from ancilla dephasing errors, but the passive scheme has a built-in advantage. 
\medskip
\subsubsection{Imperfect unitary recovery}

Both the active and passive schemes have imperfections in the unitary operation performed for recovery. For the passive case, this imperfection stems from the jump and no-jump evolution terms in the Lindblad master equation. The stochastic nature of these terms will cause a superposition of $\ket{0_L}=\frac{1}{\sqrt2}(|1\rangle+|5\rangle)$ to lose coherence and trend toward a mixed state even with a perfect correction operator. This is because the no jump and jump evolutions drive the state into different components of $|1\rangle+|5\rangle$ and $|1\rangle-|5\rangle$ in a stochastic way, while they act differently on the other code state, $\ket{1_L}=|3\rangle$, since there is no complement state. This causes a loss of phase information as one codeword is driven to a complementary state while the other is not.  It should be noted that the logical phase information is stored in the relative phase between $\ket{0_L}$ and $\ket{1_L}$, not within the code word of $\ket{0_L}$, making the logical dephasing not as severe as one may naively expect.  Nonetheless, some logical information is lost.  The loss in fidelity from this code-state distortion effect in our experiment is approximately 0.5\% per QEC cycle ($\tau_{cor}$) based on numerical calculations. 

In the active correction scheme, as in \cite{ni2023beating}, since the binomial state is fully corrected with an applied unitary accounting for jump and no jump evolution after a correction cycle, errors incurred from the correction cycle are due to the correction infidelity. There are two correction cases, correcting the no-jump evolution from no error or correcting the jump error depending on the measurement outcome. Each one of these has an infidelity associated with it, $\epsilon_{nj}$ and $\epsilon_{j}$, respectively. We use the values of $\epsilon_{nj}=1.4\%$ and $\epsilon_{j}=3.9\%$ from \cite{ni2023beating} to calculate the error rate in Table~\ref{tab:error_budget_comparison}. 

\subsubsection{Error rate comparison conclusion}

The error rates tabulated in Table~\ref{tab:error_budget_comparison} show a performance advantage for the passive approach for the $\bar{n}=3$ case.  In the active case, the QEC performance can be improved by using the smallest binomial encoding (the $\bar{n}=2$ kitten code) and reduce the cadence of QEC cycles, which then yields overall performance very similar to the passive QEC scheme demonstrated in our experiment.

\begin{figure*} % Do not use \begin{figure*}
	\centering
	\includegraphics[width=0.7\textwidth]{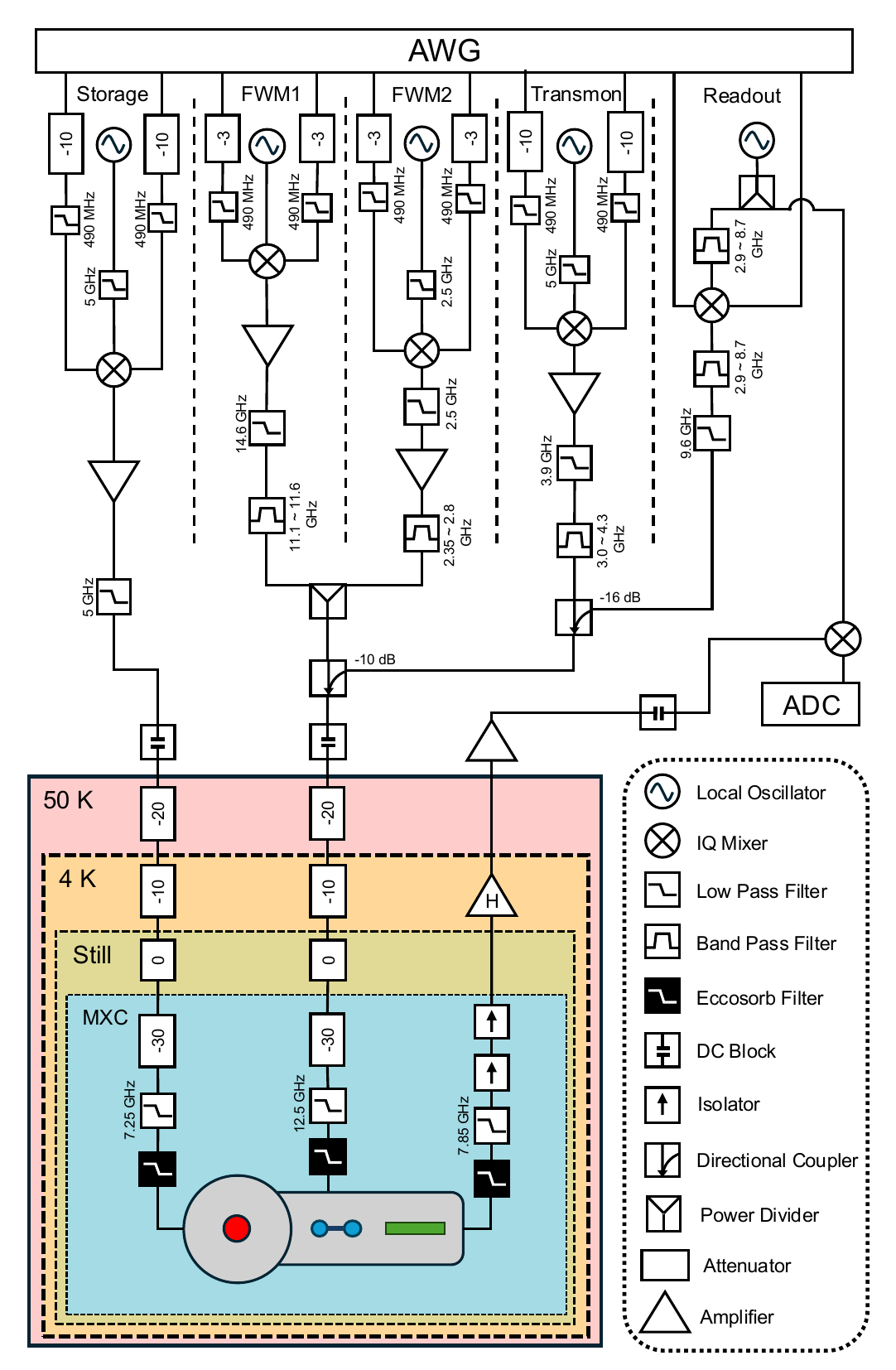} 

	% Captions go below figures
	\caption{\textbf{Dilution Fridge Diagram and Room Temperature Microwave Setup.} Filtering and attenuation/amplification setup for the experiment.
	There are five independent microwave  channels to drive the three modes in the system and two FWM processes. All the coherent control signals are generated by IQ modulation. The storage mode has a dedicated drive port. The transmon and readout modes are driven using the same input port along with the two FWM drives, that are combined using power dividers. The output port carries signal that is amplified with low-noise HEMT at 4K and room temperature amplifier at the top of the fridge. We use a combination of filters and amplifiers from mini-circuits at room temperature and marki filters suited for cryogenic purposes inside the dilution fridge. The Aluminum package containing the sapphire chip and storage cavity is housed inside a magnetic field protecting (amuneal) can that is placed on the mixing chamber of dilution fridge. }
	\label{fig:fridge_diagram} % give each figure a logical label name
\end{figure*}

\end{document}